\documentclass[a4paper,11pt]{JHEP3}

\usepackage[T1]{fontenc}
\usepackage{ae,aecompl,aeguill}		
\usepackage[ansinew]{inputenc}
\usepackage{amsmath}
\usepackage{amssymb}
\usepackage{slashed}
\usepackage{graphicx}
\usepackage{subfigure}

\DeclareTextSymbol{\degre}{OT1}{23}

\newcommand{\be}{\begin{equation}}
\newcommand{\ee}{\end{equation}}
\newcommand{\beqa}{\begin{eqnarray}}
\newcommand{\eeqa}{\end{eqnarray}}

\newcommand\CLASS{{\tt CLASS}}
\newcommand\GADGET{{\tt GADGET}}
\newcommand\CAMB{{\tt CAMB}}

\newcommand\HALOFIT{{\tt HALOFIT}}

\newcommand{\xtau}{\mathbf{x},\tau}
\newcommand{\xptau}{\mathbf{x},\mathbf{p},\tau}
\newcommand{\hub}{\mathcal{H}(\tau)}
\newcommand{\ktau}{\mathbf{k},\tau}
\newcommand{\ptau}{\mathbf{p},\tau}
\newcommand{\qtau}{\mathbf{q},\tau}

\newcommand{\keta}{\mathbf{k},\eta}
\newcommand{\peta}{\mathbf{p},\eta}
\newcommand{\qeta}{\mathbf{q},\eta}
\newcommand{\pq}{\mathbf{p},\mathbf{q}}
\newcommand{\kpq}{\mathbf{k},\mathbf{p},\mathbf{q}}
\newcommand{\kqp}{\mathbf{k},\mathbf{q},\mathbf{p}}
\newcommand{\abc}{<\varphi_a\varphi_b\varphi_c>}
\newcommand{\dbc}{<\varphi_d\varphi_b\varphi_c>}
\newcommand{\adc}{<\varphi_a\varphi_d\varphi_c>}
\newcommand{\abd}{<\varphi_a\varphi_b\varphi_d>}
\newcommand{\acd}{<\varphi_a\varphi_c\varphi_d>}
\newcommand{\cdb}{<\varphi_c\varphi_d\varphi_b>}
\newcommand{\abcd}{<\varphi_a\varphi_b\varphi_c\varphi_d>}
\newcommand{\debc}{<\varphi_d\varphi_e\varphi_b\varphi_c>}
\newcommand{\adec}{<\varphi_a\varphi_d\varphi_e\varphi_c>}
\newcommand{\abde}{<\varphi_a\varphi_b\varphi_d\varphi_e>}
\newcommand{\deltakq}{\delta_D(\mathbf{k}+\mathbf{q})}
\newcommand{\deltapj}{\delta_D(\mathbf{p}+\mathbf{j})}
\newcommand{\deltakp}{\delta_D(\mathbf{k}+\mathbf{p})}
\newcommand{\deltaqj}{\delta_D(\mathbf{q}+\mathbf{j})}
\newcommand{\deltakj}{\delta_D(\mathbf{k}+\mathbf{j})}
\newcommand{\deltaqp}{\delta_D(\mathbf{q}+\mathbf{p})}
\newcommand{\deltakpqj}{\delta_D(\mathbf{k}+\mathbf{p}+\mathbf{q}+\mathbf{j})}

\newcommand{\Hub}{\mathcal{H}}

\preprint{CERN-PH-TH/2011-097, LAPTH-014/11} 

\title{Non-linear matter
  power spectrum from Time Renormalisation Group: efficient computation and comparison with one-loop}

\author{Benjamin Audren$^{a}$, Julien Lesgourgues$^{a,b,c}$\vspace{.2cm}\\
  {$^a$}Institut de Th\'eorie des Ph\'enom\`enes Physiques,\\ \'Ecole
  Polytechnique F\'ed\'erale de Lausanne,\\ CH-1015, Lausanne,
  Switzerland.\vspace{.2cm}\\ {$^b$} CERN, Theory Division,\\ CH-1211
  Geneva 23, Switzerland.\vspace{.2cm}\\ {$^c$} LAPTh (CNRS -
  Universit\'e de Savoie), BP 110,\\ F-74941 Annecy-le-Vieux Cedex,
  France.\vspace{.2cm}}

\abstract{We address the issue of computing the non-linear matter
  power spectrum on mildly non-linear scales with efficient
  semi-analytic methods. We implemented M.~Pietroni's Time
  Renormalization Group (TRG) method and its Dynamical 1-Loop (D1L)
  limit in a numerical module for the new Boltzmann code \CLASS{}. Our
  publicly released module is valid for $\Lambda$CDM models, and
  optimized in such a way to run in less than a minute for D1L, or in
  one hour (divided by number of nodes) for TRG.  A careful comparison
  of the D1L, TRG and Standard 1-Loop approaches reveals that 
  results depend crucially on the assumed initial bispectrum at high redshift.
  When starting from a common assumption, the three methods give roughly the same results, showing that the partial resumation of diagrams beyond one loop in the TRG method improves one-loop results by a negligible amount. A comparison with  highly accurate simulations by M.~Sato \& T.~Matsubara shows that all three methods tend to over-predict non-linear corrections by the same amount on small wavelengths.
Percent precision is achieved until $k\sim
  0.2\,h$Mpc$^{-1}$ for $z \geq 2$, or until $k\sim 0.14\,h$Mpc$^{-1}$ at $z=1$.}

\begin{document} 

\section{Motivations}

Large scale structures in our universe have formed during the matter
dominated era, starting from a very homogeneous state. In order to
explain this mechanism, one must compute the evolution of the
dominating species during this period: Cold Dark Matter (CDM) and
baryons in the standard $\Lambda$CDM Model. It is generally
agreed that structures formed from the gravitational collapse of
small density perturbations. Depending on their wavelength, they
entered at different times within the Hubble scale, which plays the
role of a causal horizon for this process.
Perturbations with very large wavelength, or very small wave number,
had no time to evolve significantly beyond the linear regime. In
practice, one can apply the linear perturbation theory for wave
numbers smaller than $0.1h$Mpc$^{-1}$. In the opposite limit, for
scales under $\sim 10$~Mpc today, the presence of highly non-linear
structures such as galaxies points out the need for completely
non-linear computations. This is taken care of by N-body simulations,
that follow the evolution of a large ensemble of galaxies. 

Current and upcoming surveys such as the Sloan Digital Sky Survey~\footnote{\tt http://www.sdss.org}, 
the Large Synoptic Survey Telescope~\footnote{\tt http://www.lsst.org} and other
Large Scale Structure (LSS) experiments probe this evolution with
increasingly high precision. Through these new observations, cosmology
opens a new window to confirm, constrain or infirm different high
energy physics scenarios, related for instance to: neutrino masses,
inflationary and dark energy models, or modifications of gravity.

The discriminating power of LSS observations depends crucially on the
maximum wavenumber used in the comparison with the theory. By limiting
the analysis to linear scales with $k<k_{max}=0.1h$Mpc$^{-1}$, one
loses a lot of sensitivity, since the total amount of information
scales like $k_{max}^3$. For instance, the strong dependence of
neutrino mass error bars on $k_{max}$ is illustrated in
\cite{Hannestad:2002cn}. To further enhance the sensitivity, one would
be tempted to use highly detailed N-body simulations. Unfortunately,
in order to find both the best-fitting values and the error bars of the free
cosmological parameters of a given scenario, one needs to compute a
huge number of theoretical spectra corresponding to different points
in parameter space. With the most efficient techniques (Monte Carlo
Markov Chains), a minimum of 10'000 to 100'000 points is
necessary, depending on the complexity of the model. N-body
simulations are far too slow for being carried in each of these
points, or even in sizable fraction of them. This raises the need for
semi-analytical tools to make accurate predictions on interesting
scales. Even if such tools remain accurate only
in a small range of mildly non-linear scales (just above
$0.1h$Mpc$^{-1}$), they can play a crucial role in measuring
quantities like neutrino masses.

Some of these semi-analytical tools just consist in fitting formulas
calibrated to N-body simulation, as for instance \HALOFIT{}
\cite{Smith:2002dz}. \HALOFIT{} represents the fastest thinkable way to account
for non-linear perturbations, but its range of validity is limited to
the minimal $\Lambda$CDM model.  Extending it to non-minimal
cosmological scenarios requires many N-body simulations (and in some
cases, like for models containing species with a sizable
free-streaming length, carrying any N-body simulation remains very
involved). Moreover, \HALOFIT{} does not provide a good fit
to the Baryon Acoustic Oscillations (BAOs) in the non-linear spectrum.

Other semi-analytical methods have been proposed to actually calculate
the non-linear power spectrum in Fourier space, taking into account
the effects of mode coupling to some extent (for reviews and
comparisons, see e.g. \cite{Bernardeau01,Crocce:2005xy,Crocce:2007dt,Pietroni08,Carlson:2009it,Sato:2011qr}).  Of course, all these approaches fail when
dealing with the highly non-linear regime. However, their formulation
stays consistent within the mildly non-linear regime, up to some
$k_{max}$ which depends on the method. Any tool implementing a good
compromise between computing time on the one hand, and accuracy
(i.e. large $k_{max}$) on the other hand, can be extremely useful for
two purposes: first, for calibrating fitting formulas in the mildly
non-linear regime for extended cosmological scenarios, without running N-body
simulations; second, if the method is fast enough, for being employed
at each point in parameter space when fitting cosmological models to
the data.

In this paper, we concentrate on the Time Renormalization Group (TRG)
method proposed by Massimo~Pietroni \cite{Pietroni08}. This method
consists in integrating over time a coupled system of differential equations for the density and velocity power spectra and bispectra. Since the non-linearity computed at a given time step affect the time-derivative of the spectra at this step, the TRG method continuously includes higher-order corrections to the linear power spectra, and can be seen as a simple way to resum a sub-class of diagrams beyond one loop. 

A simple variant of the TRG equations, consisting in using products of the linear power spectra in the non-linear source term of the equations, provides strictly one-loop
results. However, the one-loop power spectrum computed in that way is
obtained from dynamical equations (we actually call this method D1L
for Dynamical 1-Loop). This should be contrasted with the Standard
1-Loop (S1L) method in which the time evolution is integrated away,
using some simplifying assumptions. The main focus of this paper
consists in a detailed comparison between S1L, D1L and TRG results for
realistic $\Lambda$CDM models. We will see that assumptions
concerning the initial bispectrum at high redshift crucially affect the results al low redshift.
This point had been overlooked in the past, and will lead us to the conclusion that when starting from the same assumptions, the three methods only differ by a negligible amount. Hence, using the TRG method (at the order discussed in the current literature) instead of the much faster D1L algorithm does not appear to be justified.

We also provide some details about our numerical implementation
of these methods in the form of a C module for the new Boltzmann code
\CLASS{}\footnote{\tt http://class-code.net}
\cite{Lesgourgues:2011re,Blas:2011rf}, released with the version v1.2
of the code.  
Our implementation allows for a small
computing time (for the TRG method, 70 minutes on a single-core CPU,
to be divided by the number of cores in a parallel execution; and for
D1L, less than a minute), opening the possibility to
integrate it within a parameter extraction code. These approaches
could easily be used e.g. for computing CMB lensing or cosmic shear
power spectra in the mildly non-linear regime, without recurring to
any N-body simulation or empirical fitting formulas. They also provide
the non-linear velocity and cross density-velocity power spectra.

We review the standard one-loop approach very briefly in
section~\ref{sec:s1l}, M.~Pietroni's TRG method in
section~\ref{sec:exact}, and our numerical implementation of the
previous equations in section~\ref{sec:num}.  We present some
self-consistency checks in the exact Einstein-De-Sitter (EdS) limit in
\ref{sec:eds}, and show our results for a realistic $\Lambda$CDM model
in section~\ref{sec:results}. We include a comparison with N-body
simulations performed with \GADGET{}~\cite{Springel:2005mi} by Sato \&
Matsubara~\cite{Sato:2011qr}, and with \HALOFIT{}.  A summary is
provided in section~\ref{sec:discussion}. Appendix \ref{sec:boltz}
recalls how the continuity and Euler equations are derived from the
Boltzmann equation.  Finally, Appendix \ref{sec:class} provides
details on the code itself.

\section{Standard one-loop approach\label{sec:s1l}}

The one-loop density power spectrum is usually computed at any
redshift using the formula
\begin{align}
  & P_\mathrm{NL}(k) =P_\mathrm{L}(k) \label{one-loop}\\
  & + \frac{2 \pi}{k} \int_0^\infty dp P_\mathrm{L}(p) \left\{ \int_{|p-\frac{k}{2}|+\frac{k}{2}}^{k+p} dq P_\mathrm{L}(q)
  \left( \frac{\left(2 k^4 + 3 k^2(q^2+p^2) + 10 p^2 q^2 - 5 (p^4 +
        q^4)\right)^2}{14^2 p^3 q^3}\right) \right. \nonumber\\
& + \left. \frac{k^3 P_\mathrm{L}(k)}{252} \left( \frac{12 k^2}{p^2} - 158 + \frac{100p^2}{k^2} - \frac{42 p^4}{k^4} - \frac{3 k^3}{p^3} \left(\frac{7 p^2}{k^2} + 2 \right)\left(\frac{p^2}{k^2}-1\right)^3 \ln \frac{\left| 1-\frac{p}{k} \right|}{1+\frac{p}{k}}\right)\right\}~, \nonumber
\end{align}
where $P_\mathrm{L}(k)$ is the linear density power spectrum (see
e.g.~\cite{Bernardeau01} and references therein). This result, that we
will call S1L for Standard 1-Loop, is obtained by expanding the
density and velocity fields describing a single-flow pressureless
fluid at order two in perturbations. It is based however on additional
approximations concerning the time-evolution of linear and non-linear
perturbations.

Indeed, the above formula corresponds to the exact one-loop solution
of the continuity and Euler equations for such a fluid only in the
perfect Einstein-De-Sitter limit (more precisely, assuming that all
modes obey to a newtonian evolution in a fully matter-dominated
universe). In this case, the matter density perturbations can be
expanded as
\begin{equation}
\delta(\tau, k) = \sum_{i=1}^{\infty} a(\tau)^i \delta^{(i)}(k)
\end{equation}
where $\tau$ stands for conformal time, $a^i$ for the scale factor 
to the power $i$, and
$\delta^{(i)}$ for the $i$-th order perturbation rescaled by $a^i$.  A
similar expansion is performed for the velocity divergence field $\theta$: 
\begin{equation}
\theta(\tau, k) = {\cal H}(\tau) \sum_{i=1}^{\infty} a(\tau)^i \theta^{(i)}(k)~,
\end{equation}
with ${\cal H}\equiv a'/a$.  When writing the relations between the
$\delta^{(i)}$ and $\theta^{(i)}$ functions, all time-dependent
factors simplify, thanks to the structure of the non-linear continuity
and Euler equations. After collecting all terms, one can write the
non-linear power spectrum as a function of the linear one evaluated at
the same redshift, without any integral over time.

In a realistic $\Lambda$CDM universe, the linear density fluctuations
undergo a complicated evolution at high redshift: their behavior is
radically different on super/sub-Hubble scales, during
radiation/matter domination, before/after photon decoupling. After
recombination, when radiation perturbations become negligible, their
evolution can be caught by a scale-independent growth factor often
parametrized as $a(\tau) \times g(\tau)$, where $g$ is a function of
time falling below one close to $\Lambda$ or Dark Energy
domination.  Similarly, the linear velocity divergence $\theta$
scales like $a(\tau) \times f(\tau)$, with $f \equiv g+a\,g'/a'$.  We
show $g(z)$ and $f(z)$ for typical $\Lambda$CDM parameters in
figure~\ref{fig:gf}.
\FIGURE{
\includegraphics[scale=0.7]{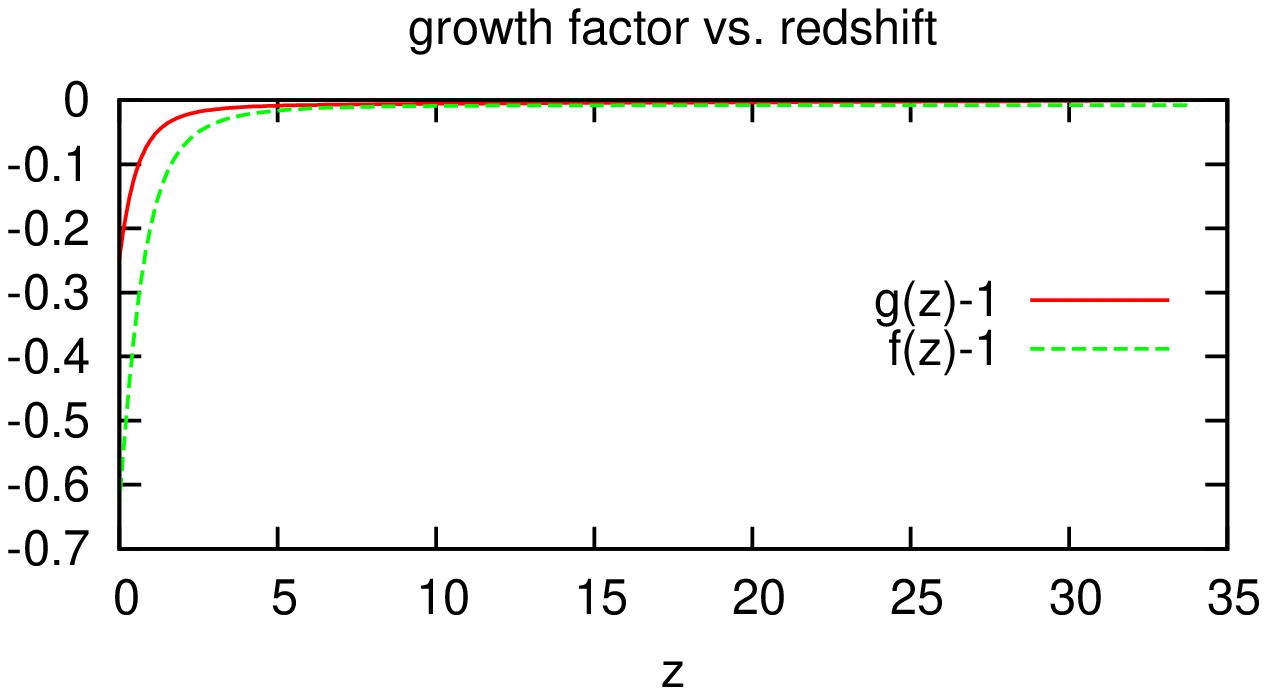}\\
\caption{\label{fig:gf}Functions $g(z)$ and $f(z)$ related to the
  linear growth factor of density perturbations and velocity
  divergence, for a $\Lambda$CDM model with
  $\Omega_{\Lambda}=0.735$.}
}
As long as one can neglect massive neutrinos or eventual dark energy
perturbations, high-order terms in perturbation theory should
still be separable functions of time and wavenumber, because the model
contains no characteristic length that could lead to a scale-dependent growth
factor within the Hubble radius. However, these terms do not
scale like powers of $[a \, g]$. This can be checked by trying an
expansion of the type
\begin{equation}
\delta(\tau, k) = \sum_{i=1}^{\infty} [a(\tau) g(\tau)]^i \delta^{(i)}(k)~,
\qquad
\theta(\tau, k) = {\cal H}(\tau) \sum_{i=1}^{\infty} [a(\tau) g(\tau)]^i \theta^{(i)}(k).
\label{eq:s1linlambda}
\end{equation}
Since the non-linear equations of motion still contain factors in
$a'/a$, not $[ag]'/[ag]$, the time-dependence cannot be factored out
in the same way as in the Einstein-De-Sitter case (for more details see e.g.
\cite{Bernardeau01,Pietroni08}). In summary, the
formula~(\ref{one-loop}) is only approximate, due to the non-trivial
evolution of perturbation in a realistic $\Lambda$CDM universe,
both at high redshift (recombination time and before) and at low redshift
($\Lambda$ domination).

In order to evaluate the corresponding error, one should compare the
power spectrum obtained from eq.~(\ref{one-loop}) with that obtained
from a full integration over time of one-loop equations. This task
becomes easy once the TRG equations have been correctly implemented,
since the TRG method relies on dynamical equations, and since the
one-loop order corresponds to a simple limit of these equations. A
test presented in \cite{Pietroni08} suggests that the error induced by
low-redshift effects is very small. Below, we will confirm this
finding, but interestingly, we will show that the error induced by
mistreating the initial bispectrum at high redshift
is instead very significant.

\section{Time Renormalization Group equations\label{sec:exact}}

For the rest of this article, the following conventions are
assumed. The universe is considered to be described by the
$\Lambda$CDM model, with zero spatial curvature. 
We assume Gaussian initial
conditions, and therefore a vanishing bispectrum at initial time.

During the matter dominated era and within the Hubble radius, the
matter velocity dispersion is small. It is therefore possible to use
the Newtonian limit of the Einstein's field equations for the
perturbation of the energy tensor. By starting TRG computations at a
redshift $z_{ini}=35$, and dealing only with non-linear equations for
wavenumbers above $2.10^{-3}~h/Mpc$, one stays in the range where
these approximations are valid.

In the TRG approach and within the $\Lambda$CDM paradigm, one has to
renormalize both baryon and CDM. This could lead to a tedious system
with mixed variables, since \textit{a priori} baryons have
non-negligible pressure (sound speed). However, pressure effects
appears at very small scales (in any case, above $10~h/\text{Mpc}$),
and it is safe to approximate baryons as pressureless and
collisionless. Then, both baryon and CDM follow the same non-linear
equations, and share the same initial condition provided that the
starting redshift is chosen sufficiently far from decoupling (which is
the case with $z_{ini} = 35$). They can be described as a single
perturbation field, hereafter denoted as the matter field with
subscript $m$, and all quantities $X_m$ are defined as
$\Omega_mX_m=\Omega_bX_b+\Omega_{c}X_{c}$.

The non-linear collisionless Boltzmann equation of a self-gravitating
fluid is the master equation describing structure formation, and being
solved in N-body simulations. In Appendix A, we recall the basic steps
and assumptions allowing to reduce the Boltzmann equation to a coupled
system of non-linear equations in Fourier space: namely, the continuity and
Euler equation for density perturbations $\delta_m(\ktau)$ and velocity
divergence $\theta_m(\ktau)$.

In order to solve these equations, the most common approach consists
in expanding $\delta_m$ and $\theta_m$ in perturbations, and solve the
system at a given order. The TRG relies on a different
expansion, in connected $n$-point correlation functions. The two approaches
start however from  the same equations. One needs to introduce simplified
notations similar to those in \cite{Crocce:2005xy}, and to define a doublet
$\varphi_a$ (a=1,2):
\begin{align}
  \begin{pmatrix}\varphi_1(\keta) \\ \varphi_2(\keta)\end{pmatrix}\equiv e^{-\eta}\begin{pmatrix}\delta_m(\keta)\\ -\theta_m(\keta)/\Hub\end{pmatrix}~,
  \label{eqn:doubletdef}
\end{align}
where $\eta=\log(a/a_{\text{ini}})$. The role of the exponential term
consists in factorizing out most of the time evolution. Indeed, the
variables $\varphi_i(\keta)$ are constant during matter domination and
in the linear regime; their subsequent slow variation is easier to
catch numerically than the fast variation of $(\delta_m, \theta_m)$.
The non-linear continuity and Euler equations 
(eqs.~(\ref{eqn:contF},\ref{eqn:eulerF}) of Appendix A) can then be
cast in a compact form:
\begin{align}
  \partial_{\eta}\varphi_a(\keta)=-\Omega_{ab}(\keta)\varphi_b(\keta)+e^{\eta}\gamma_{abc}(\mathbf{k},-\mathbf{p},-\mathbf{q})\varphi_b(\peta)\varphi_c(\qeta),\label{eqn:compact}
\end{align}
where we follow the Einstein convention for repeated indices (sum) and
momentum (integral on $\mathbf{p}$ and $\mathbf{q}$).
The vertex functions $\gamma_{abc}(\kpq)$
(a,b,c=1,2), encoding the non-linearity, are defined through:
\begin{align}
  \gamma_{121}(\kpq)&=\frac12\delta_D(\mathbf{k}+\mathbf{p}+\mathbf{q})\alpha(\mathbf{p},\mathbf{q}),\label{eqn:defgamma121}\\
  \gamma_{222}(\kpq)&=\delta_D(\mathbf{k}+\mathbf{p}+\mathbf{q})\beta(\mathbf{p},\mathbf{q}),\label{eqn:defgamma222}\\
  \gamma_{112}(\kpq)&=\gamma_{121}(\kqp),\label{eqn:defgamma112}
\end{align}
and all other components
vanish.  The definition of the kernels $\alpha$ and $\beta$ is given in
eqs.~(\ref{eqn:alphabeta}) of Appendix~\ref{sec:boltz}. The $\Omega$ matrix, accounting
for the linear evolution, reads
\begin{align}
\begin{pmatrix}
    1 & -1 \\ 
    -\frac32\Omega_m(\eta)(1+\mathcal{B}(\keta)) & 2+\frac{\hub '}{\hub}
\end{pmatrix},\label{eqn:defOmega}
\end{align}
including a factor $\mathcal{B}(\keta)=\sum_{i\neq m}
\frac{\Omega_i\delta_i}{\Omega_m\delta_m}$ (in $\Lambda$CDM, $i$
runs only on photon and neutrino species) which appeared thought the Poisson
equation.  Although we are interested in following the matter
perturbations non-linearly, we do not want to abandon altogether other
species which might play an important role at the percent
precision. The equations presented in this section include this
effect, while those solved by the code do not. This issue will be
discussed in more details in sections \ref{sec:num} and
\ref{sec:discussion}.
%
%
%
A nice
property of this formulation is that everything one would like to test about
new cosmologies is entirely encoded in the $\Omega$ matrix, while the
non-linear vertices are model-independent.

The equations of evolution for correlation functions of $\varphi_a$
can be derived by applying several times equation (\ref{eqn:compact}),
omitting arguments for clarity:
\begin{align}
  &\partial_{\eta}<\varphi_a\varphi_b> & = &-\Omega_{ac}<\varphi_c\varphi_b> - \Omega_{bc}<\varphi_a\varphi_c>\nonumber\\
  &&&+e^{\eta}\gamma_{acd}\cdb+e^{\eta}\gamma_{bcd}\acd,\label{eqn:dpab_deta}\\
  &\partial_{\eta}\abc & = &-\Omega_{ad}\dbc - \Omega_{bd}\adc - \Omega_{cd}\abd\nonumber\\
  &&&+e^{\eta}\gamma_{ade}\debc+e^{\eta}\gamma_{bde}\adec\nonumber\\
  &&&+e^{\eta}\gamma_{cde}\abde,\label{eqn:dpabc_deta}\\
  &\partial_{\eta}\abcd & = & \dots \nonumber
\end{align}
The power spectrum $P_{ab}$, bispectrum $B_{abc}$ and trispectrum
$Q_{abcd}$ of the doublet components are defined as:
\begin{align}
  &<\varphi_a(\keta)\varphi_b(\qeta)> ~\equiv ~\delta_D(\mathbf{k}+\mathbf{q})P_{ab}(\keta)\nonumber\\
  &<\varphi_a(\keta)\varphi_b(\qeta)\varphi_c(\peta)> ~\equiv ~ \delta_D(\mathbf{k}+\mathbf{q}+\mathbf{p})B_{abc}(\mathbf{k},\mathbf{q},\mathbf{p};\eta)\nonumber\\
  &<\varphi_a(\keta)\varphi_b(\qeta)\varphi_c(\peta)\varphi_d(\mathbf{j},\eta)>~\equiv\nonumber\\
  &\hspace{2cm} \left[ \deltakq\deltapj P_{ab}(\keta)P_{cd}(\peta)\right.\nonumber\\
  &\hspace{2cm} + \deltakp\deltaqj P_{ac}(\keta)P_{bd}(\qeta)\nonumber\\
  &\hspace{2cm} + \deltakj\deltaqp P_{ad}(\keta)P_{bc}(\qeta)\nonumber\\
  &\hspace{2cm}\left. + \deltakpqj Q_{abcd}(\mathbf{k},\mathbf{q},\mathbf{p},\mathbf{j};\eta)\right]\label{eqn:def_spectra}
\end{align}
One has to note that $P_{ab}$ will differ from the usual
matter/velocity power spectrum by a factor $e^{2\eta}$ and some
factors of $\hub$ in the case of $P_{\delta\theta}$ and
$P_{\theta\theta}$. To close this system of equations, the
approximation proposed by M.~Pietroni consists in setting $Q_{abcd}$ to
zero. Hence, the TRG does not consist in truncating at a given order
in perturbations or loops, but at a given order in connected
correlation functions.  Assuming $Q_{abcd}=0$ should be valid up to
some time in the non-linear evolution. This truncation still allows
for a departure from gaussianity (the bispectrum is not set to zero at
any time), and for evolving models with a non-zero primordial
bispectrum.  By putting together (\ref{eqn:def_spectra}) and
(\ref{eqn:dpab_deta}, \ref{eqn:dpabc_deta}), one finds a closed system
of equations, corresponding to the TRG equations truncated at the
trispectrum order:
\begin{align}
  \partial_{\eta}P_{ab}(\keta)=&-\Omega_{ac}(\keta)P_{cb}(\keta)-\Omega_{bc}(\keta)P_{ac}(\keta)\nonumber\\
  &+e^{\eta}\int \text{d}^3\mathbf{q}\left[\gamma_{acd}(\mathbf{k},-\mathbf{q},\mathbf{q}-\mathbf{k})B_{bcd}(\mathbf{k},-\mathbf{q}, \mathbf{q}-\mathbf{k};\eta)\right.\nonumber\\
  &\hspace{2cm}\left.+B_{acd}(\mathbf{k},-\mathbf{q}, \mathbf{q}-\mathbf{k};\eta)\gamma_{bcd}(\mathbf{k},-\mathbf{q}, \mathbf{q}-\mathbf{k},\eta)\right],\label{eqn:dpab}\\
  \partial_{\eta}B_{abc}(\mathbf{k},-\mathbf{q}, \mathbf{q}-\mathbf{k};\eta) = & -\Omega_{ad}(\keta)B_{dbc}(\mathbf{k},-\mathbf{q}, \mathbf{q}-\mathbf{k};\eta)\nonumber\\
  &-\Omega_{bd}(-\mathbf{q},\eta)B_{adc}(\mathbf{k},-\mathbf{q}, \mathbf{q}-\mathbf{k};\eta)\nonumber\\
  &-\Omega_{cd}(\mathbf{q}-\mathbf{k},\eta)B_{abd}(\mathbf{k},-\mathbf{q}, \mathbf{q}-\mathbf{k};\eta)\nonumber\\
  &+2e^{\eta}\left[\gamma_{ade}(\mathbf{k},-\mathbf{q}, \mathbf{q}-\mathbf{k})P_{db}(\mathbf{q},\eta)P_{ec}(\mathbf{k}-\mathbf{q},\eta)\right.\nonumber\\
  & \hspace{1.3cm}\gamma_{bde}(-\mathbf{q}, \mathbf{q}-\mathbf{k},\mathbf{k})P_{dc}(\mathbf{k}-\mathbf{q},\eta)P_{ea}(\mathbf{k},\eta)\nonumber\\
  & \hspace{1.2cm}\left.\gamma_{cde}(\mathbf{q}-\mathbf{k},\mathbf{k},-\mathbf{q})P_{da}(\keta)P_{eb}(\qeta)\right].\label{eqn:dbabc}
\end{align}
These definitions are in agreement with the Fourier transform
conventions adopted in this paper, namely, $A(\ktau)=(2\pi)^{-3}
\int\text{d}^3\mathbf{x}\exp(-i\mathbf{k}\mathbf{x})A(\xtau)$. The other common
definition, actually used in \CLASS{}, reads
$A(\ktau)=\int\text{d}^3\mathbf{x}\exp(-i\mathbf{k}\mathbf{x})A(\xtau)$.  In
this case, in the RHS of eq.~(\ref{eqn:dbabc}), the last term picks up one
extra $(2\pi)^3$ factor.

\section{Numerical implementation\label{sec:num}}

\subsection{Integrated form of the equations\label{ssec:integ}}

Our module does not rely on the full TRG equations, but on their
integrated form already introduced in the original TRG
paper~\cite{Pietroni08}, that we briefly summarize below.

As a consequence of the statistical homogeneity and isotropy of the universe,
the power spectrum $P$ only depends on $k=|\mathbf{k}|$, while the
$\gamma$ and $B$ functions that appear in (\ref{eqn:dpab}) under the
summation sign only depend of $|\mathbf{q}|$,$|\mathbf{k}-\mathbf{q}|$
and the angle between $\mathbf{k}$ and $\mathbf{q}$. Let us introduce
a spherical coordinate system $(r,\theta,\varphi)$ such that
$\mathbf{k}=(k,0,0)$ and $\mathbf{q}=(q,\theta,\varphi)$. When
integrating over $\text{d}^3 \mathbf{q}$, the integral over $\varphi$ yields a $2\pi$
factor. Then by performing the variable change between $\theta$ and
$p$ defined as $p^2=|k+q|^2$, one can transform the measure to:
\begin{align}
  \int\text{d}^3\mathbf{q}=\frac{2\pi}{k}\int_0^{\infty}\hspace{-0.3cm}q\text{d}q\int_{|q-k|}^{q+k}\hspace{-0.3cm}p\text{d}p.
  \label{eqn:measure}
\end{align}
Finally, by analyzing the shape of the integration domain, it is
possible to further simplify the integrand in (\ref{eqn:dpab}) and get
\begin{align}
  \partial_{\eta}P_{ab}(k,\eta)=&-\Omega_{ac}(k,\eta)P_{cb}(k,\eta)-\Omega_{bc}(k,\eta)P_{ac}(k,\eta)\nonumber\\
  &+e^{\eta}\frac{4\pi}{k}\int_{k/2}^{\infty}\hspace{-0.3cm}q\text{d}q\int_{|q-k|}^q\hspace{-0.3cm}p\text{d}p\left[\gamma_{acd}(k,q,p)B_{bcd}(k,q,p;\eta)\right.\nonumber\\
  &\hspace{2cm}\left.+B_{acd}(k,q,p;\eta)\gamma_{bcd}(k,q,p)\right],\label{eqn:dpab2}
\end{align}
where $\gamma_{abc}(k,q,p) \equiv
\gamma_{abc}(\mathbf{k},\mathbf{q},\mathbf{p})|_{\mathbf{p}=-(\mathbf{k}+\mathbf{q})}$. In
a numerical implementation, the bispectrum, being a function of three
variables and three indices, is a large object. To reduce the
dimensionality of the problem, it was already noticed by M.~Pietroni
that as long as the scale-dependence of the $\Omega$ matrix can be
neglected, we can obtain a closed system in which the integrated
quantities replacing the bispectrum read
\begin{align}
  I_{acd,bef}(k)&\equiv\int_{k/2}^{\infty}\hspace{-0.3cm}q\text{d}q\int_{|q-k|}^q\hspace{-0.3cm}p\text{d}p\frac12\left[ \gamma_{acd}(k,q,p)B_{efg}(k,q,p)+(q\leftrightarrow p)\right].
  \label{eqn:defI}
\end{align}
According to eq.~(\ref{eqn:dbabc}), the time evolution of $I$ is given by
\begin{align}
  \partial_{\eta}I_{acd,bef}(k)=-\Omega_{bg}I_{acd,gef}(k)
  -\Omega_{eg}I_{acd,bgf}(k) -\Omega_{fg}I_{acd,beg}(k) +
  2e^{\eta}A_{acd,bef}(k),\label{eqn:dIfriendly}
\end{align}
with the $A$'s being given by
\begin{align}
  A_{acd,bef}(k)\equiv \int_{k/2}^{\infty}dq~q\int_{|q-k|}^{q}dp~p\frac12 \left\{\gamma_{acd}(k,q,p)\left[\gamma_{bgh}(k,q,p)P_{ge}(q)P_{hf}(p) +\right.\right.\nonumber\\
  \left.\left.\gamma_{egh}(q,p,k)P_{gf}(p)P_{hb}(k)+\gamma_{fgh}(p,k,q)P_{gb}(k)P_{he}(q)\right]
  + (q\leftrightarrow p)\right\}.\label{eqn:defA}
\end{align}
In eq.~(\ref{eqn:dpab2}), all integrals have been absorbed in the definition of $I$:
\begin{align}
  \partial_{\eta}P_{ab}(k)&=-\Omega_{ac}P_{cb}(k)-\Omega_{bc}P_{ac}(k)+e^{\eta}\frac{4\pi}{k}[I_{acd,bcd}(k)+I_{bcd,acd}(k)].\label{eqn:dpabfriendly}
\end{align}
Finally, we can perform a rotation from the $(q,p)$ basis into a new $(x,y)$ basis in order to obtain
separable integration ranges:
\begin{align}
  \int_{k/2}^{\infty}dq~q&\int_{|q-k|}^{q}dp~p\left[ F(k,q,p) + (q\leftrightarrow p)\right]\nonumber\\
  &=\int_{k/\sqrt{2}}^{\infty}\text{d}x\int_0^{k/\sqrt{2}}\text{d}y\frac{x^2-y^2}{2}\left[ F(k,\frac{x+y}{\sqrt{2}},\frac{x-y}{\sqrt{2}})+ (y\leftrightarrow -y)\right],\label{eqn:varchangexy}
\end{align}
where $F$ is a generic notation for the integrands appearing in eq.~(\ref{eqn:defA}).

We implemented this integrated approach in our module as a first step
(which is already far from trivial from the numerical point of view).
Having checked this level of approximation, we aim to go back to the
full set of equations in the future. This would allow to compute the
full bispectrum evolution, starting from Gaussian or non-Gaussian
initial conditions. It would also allow to treat consistently the case
of a matrix $\Omega$ with a significant scale-dependence, and to
perform a more robust test of the approximation used in
\cite{LesgourguesPietroni09,D'Amico:2011pf,Anselmi:2011ef}, in which
the above method was employed in the presence of massive neutrinos or
dark energy perturbations.

By analyzing the symmetries of eq. (\ref{eqn:defA}), (\ref{eqn:defI})
and (\ref{eqn:defgamma121}), (\ref{eqn:defgamma222}), one can reduce
the number of independent $I_{acd,bef}(k)$ functions from $64$ to
$14$: namely, the set $(121,def)$ with
$(def)=(111),(211),(121),(112),(122),(212),(221),(222)$ and the set
$(222,ghi)$ with
$(ghi)=(111),$$(211),$$(121),$$(122),$$(212),$$(222)$.  Hence, we must
solve a system of 17 differential equations (3 for $P_{ab}$ and 14 for
$I_{acd,bef}$). The only time-consuming step is the evaluation of the
14 two-dimensional integrals of eq.~(\ref{eqn:defA}) at each time
step.

\subsection{The three modes\label{sec:3modes}}

Our numerical module can be tuned in such a way to solve either the above
equations, or some limit of these equations, in order to
reproduce different approaches. 

\subsubsection{Linear mode}

First, by setting all the $A$ coefficients to zero, we can integrate
the linear part of equations (\ref{eqn:dIfriendly},
\ref{eqn:dpabfriendly}) and try to reproduce the linear evolution of
the power spectrum between $z_{ini}$ and today. Comparing these
results with very accurate ones produced directly by the Boltzmann
code is actually a useful way to estimate the precision of the time
integration algorithm inside the TRG module, as we shall see in
section~\ref{ssec:timeint}. This calculation is performed when the
\CLASS{} input parameter ``{\tt non linear}'' is set to ``{\tt
  test-linear}''.

\subsubsection{TRG mode}

The full solution of the TRG equations, computed when the \CLASS{}
input parameter ``{\tt non linear}'' is set to ``{\tt trg}'', is of
course much more time-consuming, since $14$ independent $A$ terms must
be computed at each time step. The code has been parallelized with
{\tt OpenMP}. Its running time is of the order of 70 minutes on a
single 2.3GHz core for the set of default accuracy parameters
described in the next subsections and in Appendix \ref{sec:class}, and
on a multi-core machine the running time scales almost linearly
(i.e. 5 minutes on 14 cores, which is an optimal choice due to the
structure of the equations).

\subsubsection{Dynamical 1-Loop (D1L) mode}

Next, by computing the integrals of eq.~(\ref{eqn:defA}) using the
linear power spectra at each time, one expects to reproduce standard
results from one-loop perturbation theory.  In this paper, we will
call this method D1L for Dynamical 1-Loop.  As explained in
\cite{Pietroni08}, in an Einstein-De-Sitter universe, the D1L
equations can be integrated over time analytically, and then reduced
to eq.~(\ref{one-loop}).  In realistic cosmological models, the linear
spectra $P_{ab}$ are time dependent, and the 14 $A$ terms need
to be evaluated at each time, leading to an algorithm essentially as
time-consuming as the full TRG one.  However, as long as we stick to
$\Lambda$CDM models, we can use the fact that the linear spectra
evolution is accounted by scale-independent growth factors, in such a
way that $P_{11}^{\rm linear}(k,\eta) \propto g^2(\eta)$, $P_{22}^{\rm
  linear} \propto f^2$, and $P_{12} ^{\rm linear} \propto g f$.  So,
the time dependence of the integrals of eq.~(\ref{eqn:defA}) computed
with the linear spectra factors out, leading to a very fast algorithm.

In practise, before starting the computation, we get $g(\eta)$ and
$f(\eta)$ at each time-step using other \CLASS{} modules. We compute
the integrals of eq.~(\ref{eqn:defA}) at initial time $\eta_{\rm
  ini}$. Then, at each new time step, we source the TRG equations with
the same integrals rescaled by the appropriate number of factors
$(g/g_{\rm ini})$ and $(f/f_{\rm ini})$.  This calculation is
performed in the publicly released module when the \CLASS{} input
parameter ``{\tt non linear}'' is set to ``{\tt one-loop}''. It
provides one-loop results for $\Lambda$CDM models with no
approximation regarding the time evolution of perturbations, while
being nearly as fast as the S1L method. We will see in
section~\ref{sec:results} that it greatly improves over S1L
predictions.

\subsection{Time integration\label{ssec:timeint}}

The time integration is performed with a predictor-corrector method.
Starting from a time $\eta$ at which all quantities are known, one
computes the new $P$ and $I$ functions at the step $\eta+\delta\eta/2$
with the standard Euler method, as well as their derivatives in this
point.  The values of $P$ and $I$ at $\eta+\delta\eta$ are then
derived with the Euler method, using however the derivatives at the
mid-point.  With respect to a basic Euler method, this allows to take
many less steps for the same level of accuracy (typically, we use
between $50$ and $100$ time steps to evolve from $z=35$ to the present
time).

With a predictor-corrector method on $100$ time steps, and at all
redshifts and wave-numbers of interest, we find that the linear
spectrum calculated by the TRG module agrees at the $0.05\%$ level
with the one calculated by \CLASS{} (with many more time steps and a
sophisticated integrator).  With $50$ time steps, the error is still
as low as $0.08\%$, allowing us to keep this choice as a default
setting in the released version of the code. The convergence has also
been controlled for the one-loop computation: with only $50$ steps and
a predictor-corrector method, one reaches the same $0.1\%$ level
accuracy as with $10'000$ steps and a standard Euler method.

\FIGURE{
\includegraphics[scale=0.7]{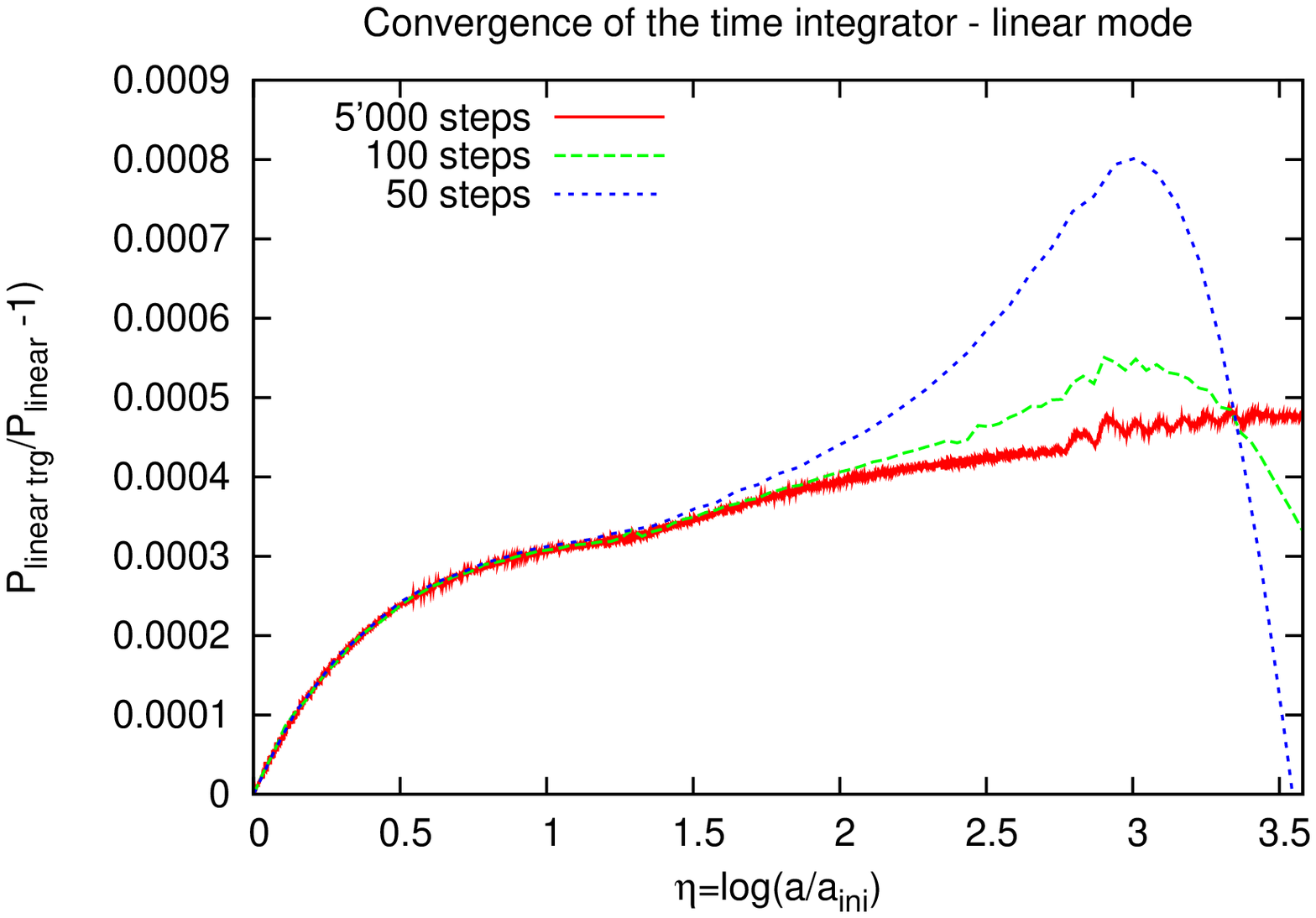}
\caption{\label{fig:conv_lin} Convergence proof of the linear
  time-evolution in the TRG module w.r.t. the linear evolution in
  \CLASS.}
}

\subsection{Momentum integration\label{ssec:mom_int}}

Since in the TRG module, most of the time is spent in the computation
of $A$ functions, an obvious way to speed up the code would be to make
a cut in the domain of integration. This domain, defined in eq. (\ref{eqn:varchangexy}),
has a rectangular shape ($x>k/\sqrt{2}$, $0<y<k/\sqrt{2}$).

After noticing that the main contribution to the integral comes from
the region near the $x=k/\sqrt{2}$ and $y=k/\sqrt{2}$ boundaries, and
that the integrand exhibits an approximate symmetry around the axis
$y=\sqrt{2}k-x$, we first tried to integrate on L-shaped bands
starting from the top left corner in $(x,y)$ space. One could hope
that after integrating over a few such L's, the integration would
converge and could be stopped. This does not work, and we noticed that
the tails of the integrand over the whole domain contribute enough to
create departures from the exact result when implementing a cut. Since
no good trade-off was found, we kept the whole domain. We use however
logarithmic integration steps in the ($x,y$) space, defined in such
way to {\it (i)} increase the resolution near the top left corner in
$(x,y)$ space and {\it (ii)} preserve the approximate axial symmetry.
Moreover, the value of these logarithmic step sizes is modulated as a
function of the wavenumber $k$.
The key parameter controlling the trade between precision and overall
speed of the module is called {\tt logstepx\_min}. This parameter
provides a lower bound on the logarithmic step used in both $x$ and
$y$ directions, for any wavenumber $k$. We show in figure
\ref{fig:conv_logstepx} that setting this parameter to 1.06 is
sufficient for the final result to converge at the 0.6\% level in the
one-loop case (at $z=1$ and for $k \leq 0.5h/$Mpc), and 0.3\% level in
the TRG case. Performing the test at $z=0$ gives similar results.

\FIGURE{
\includegraphics[scale=0.5]{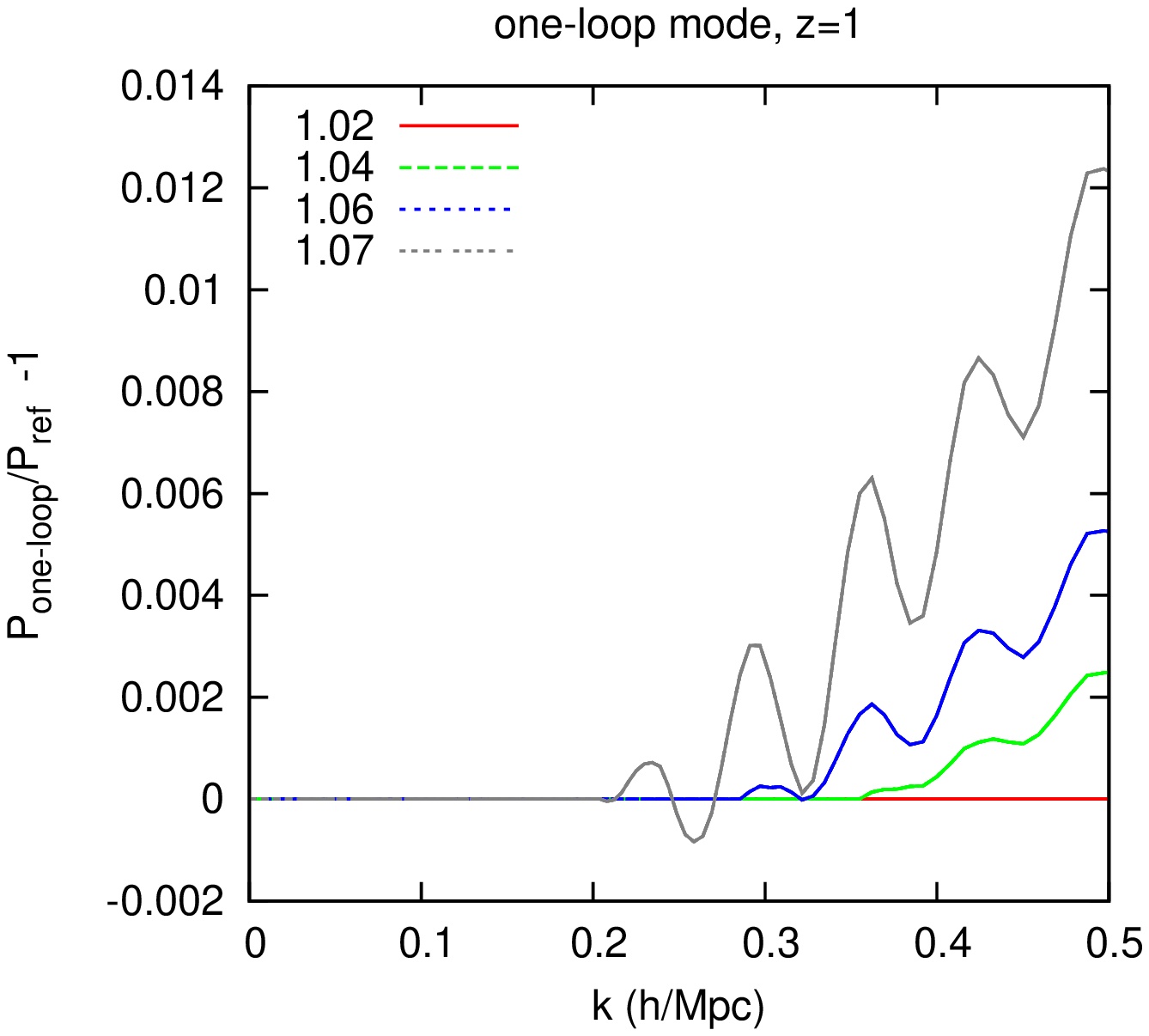}
\includegraphics[scale=0.5]{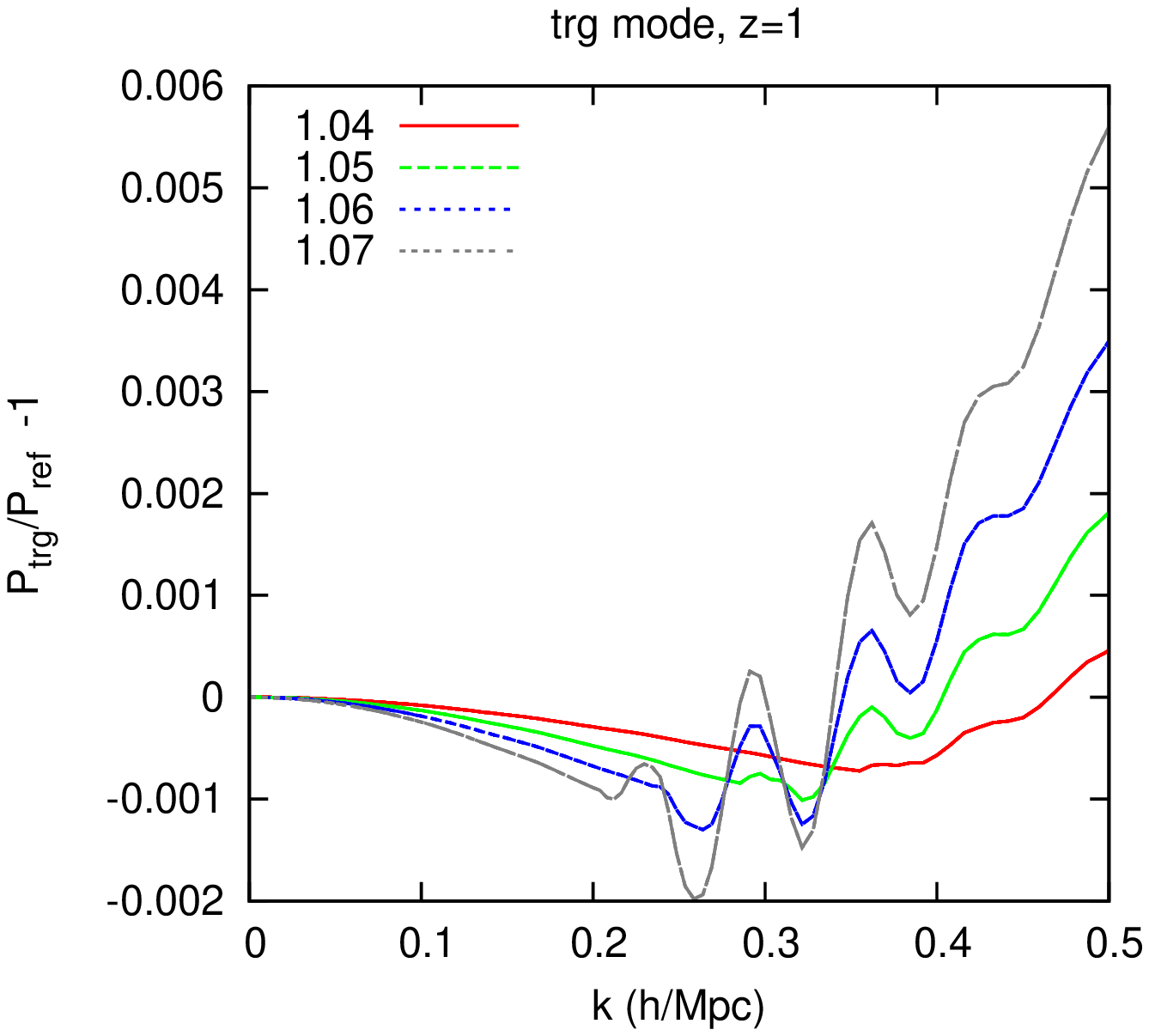}
\caption{\label{fig:conv_logstepx} Convergence proof for the integration in momentum space in the the one-loop (left) and TRG (right) modes. The reference model
has {\tt logstepx\_min}=1.001 in the one-loop case, or  {\tt logstepx\_min}=1.02 in the TRG case.}
}

In all integrals, we define some ultra-violet and infra-red cuts-off
($k_\textrm{min}$, $k_\textrm{max}$) besides which the power spectra
are assumed to vanish. Moreover, below a scale $k_L$, we assume that
the evolution of the spectrum remains strictly linear at all times:
this means that for $k<k_L$, we set the $A$ coefficients to zero, in
order to speed up the calculation.  We checked that the results are well
converged with the choices $k_\textrm{min}=10^{-4}h/$Mpc,
$k_L=10^{-3}h/$Mpc, $k_\textrm{max}=10^3h/$Mpc. 

\subsection{Numerical instabilities}

The main challenge in any implementation of the TRG algorithm is to
keep numerical noise under control in the ultra-violet limit.  Indeed,
a small error in the tail of the power spectrum would be amplified by
the recursion over time, and lead to exponential divergences. By the
same phenomenon, a slightly non-smooth initial spectrum would produce
non physical divergences. The smoothness of the spectrum is very
difficult to control since it is passed in the form of tabulated
values.

We checked that the divergences and wiggles observed in our first,
straightforward implementation of the algorithm were only numerical
artifacts. Indeed, we varied the size of time and wavenumber steps,
and observed that these features propagated at a rate defined by the
number of discrete steps, rather than by physical time.

In order to eliminate these numerical artifacts, we tried to implement
several smoothing algorithms and several exponential cut-off functions
in the small-scale linear power spectrum. Every time, some feature
would finally propagate. By the very nature of the non-linear
equations, as time goes by, any small error is amplified in huge
proportions.

We solved this issue with an approach which is physically justified by
the fact that during structure formation, mode coupling leads to a
transfer of power from large scales to small scales, and not in the
other direction. At each new time step, numerical artefact's
unavoidably affect the last four points in the list of discrete $k$
values (two points in the computation of the time derivative, plus two
points in the calculation of the $A$ integrals). Hence, one can
systematically throw away these last four points.  In practice, this
method (that we call ``double escape'') consists in integrating the
TRG equation only in a range $k_\mathrm{L} \leq k \leq
k_\mathrm{DE}(\eta)$. The wavenumber $k_\mathrm{DE}(\eta)$ (where DE
stands for double escape) coincides with $k_\mathrm{max}$ at initial
time, and then decreases by discarding four more points in $k$-space
at each half time step $\delta \eta/2$.  Since a given wavenumber is
impacted by the evolution of smaller wavenumbers only, our method
cannot affect the final results.  We checked it explicitly by
changing the step sizes in $k$ space. Such a reduction means that we
decrease $k_\mathrm{DE}(\eta)$ differently as a function of time, but
the final results remain totally invariant. Simply, smaller step sizes
allow to get results at a larger $k_\mathrm{DE}$ at $z=0$, starting
from the same initial $k_\mathrm{max}$. In the released version, the
default time steps are such that $k_\mathrm{DE}\sim 0.25h/$Mpc
today. When the user asks for non-linear spectra at a given redshift,
they are only written in output files up to the corresponding value of
$k_\mathrm{DE}(\eta)$.

Note that when computing the $A$ integrals, the power spectra still
need to be evaluated between $k_\mathrm{DE}(\eta)$ and
$k_\mathrm{max}$, but the final results have a very weak dependence on
$P_{ab}(k)$ in this range. We use a log-linear extrapolation of
$P_{ab}(k)$ besides $k_\mathrm{DE}(\eta)$, based on an estimate of the
logarithmic derivative in $k=k_\mathrm{DE}(\eta)$. We actually tried
alternative extrapolation schemes (e.g. with constant second
derivative in log-log space) and found that the final result is
independent on the scheme (again, as a consequence of the fact that a
wavenumber is only affected by its smaller neighbors).

We conclude that this ``double escape'' is a simple and robust way to
keep numerical instabilities under control, leading to a ``loss of
information'' on the smallest wavelengths, but introducing no bias in
the final results.

Note that these problems arise when dealing with the full TRG
equations.  In its one-loop limit D1L, which will be seen below to
provide excellent results, the integration over time is perfectly stable, 
and there is no need to throw points away.

\FIGURE{
\includegraphics[scale=0.5]{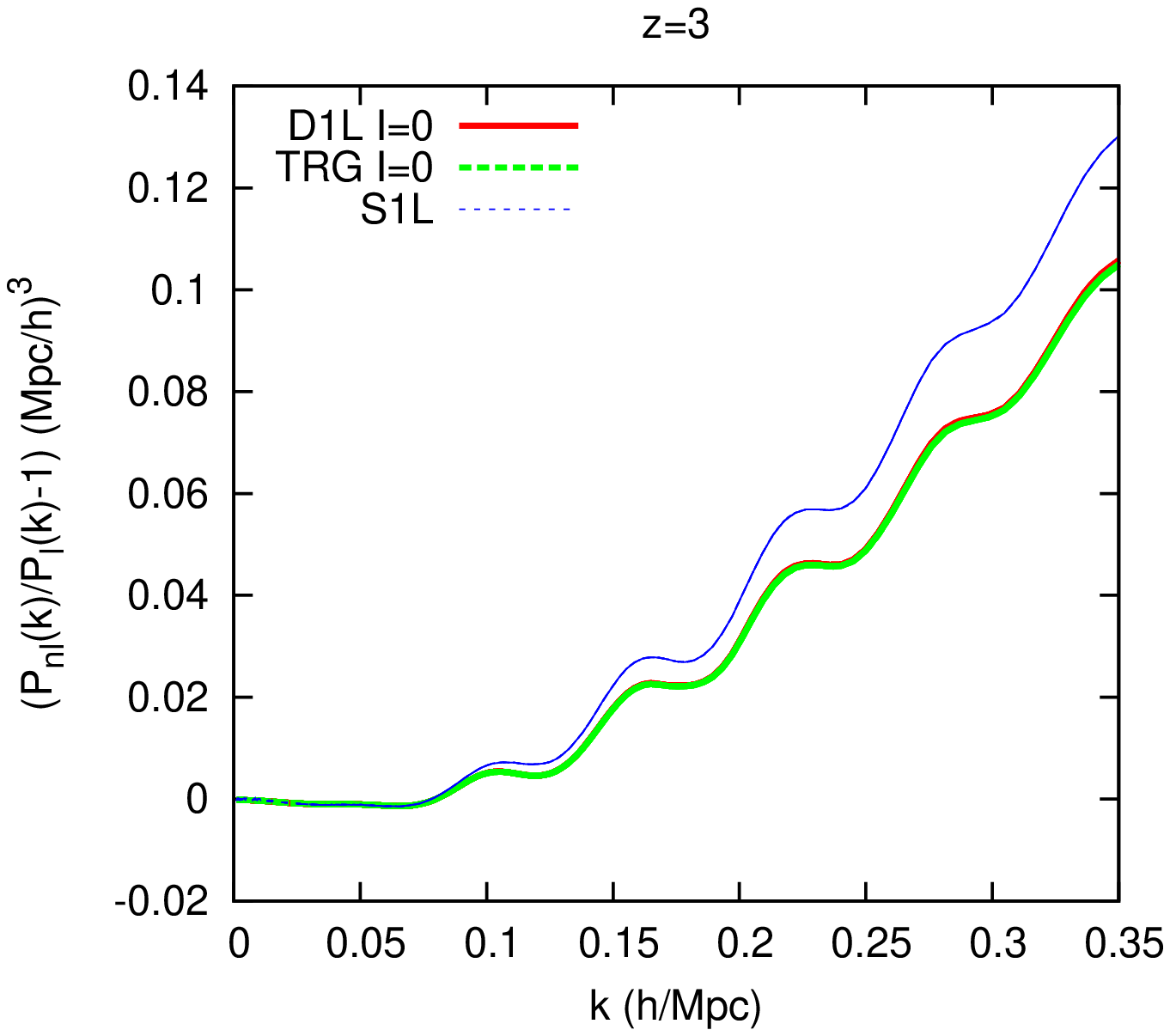}
\includegraphics[scale=0.5]{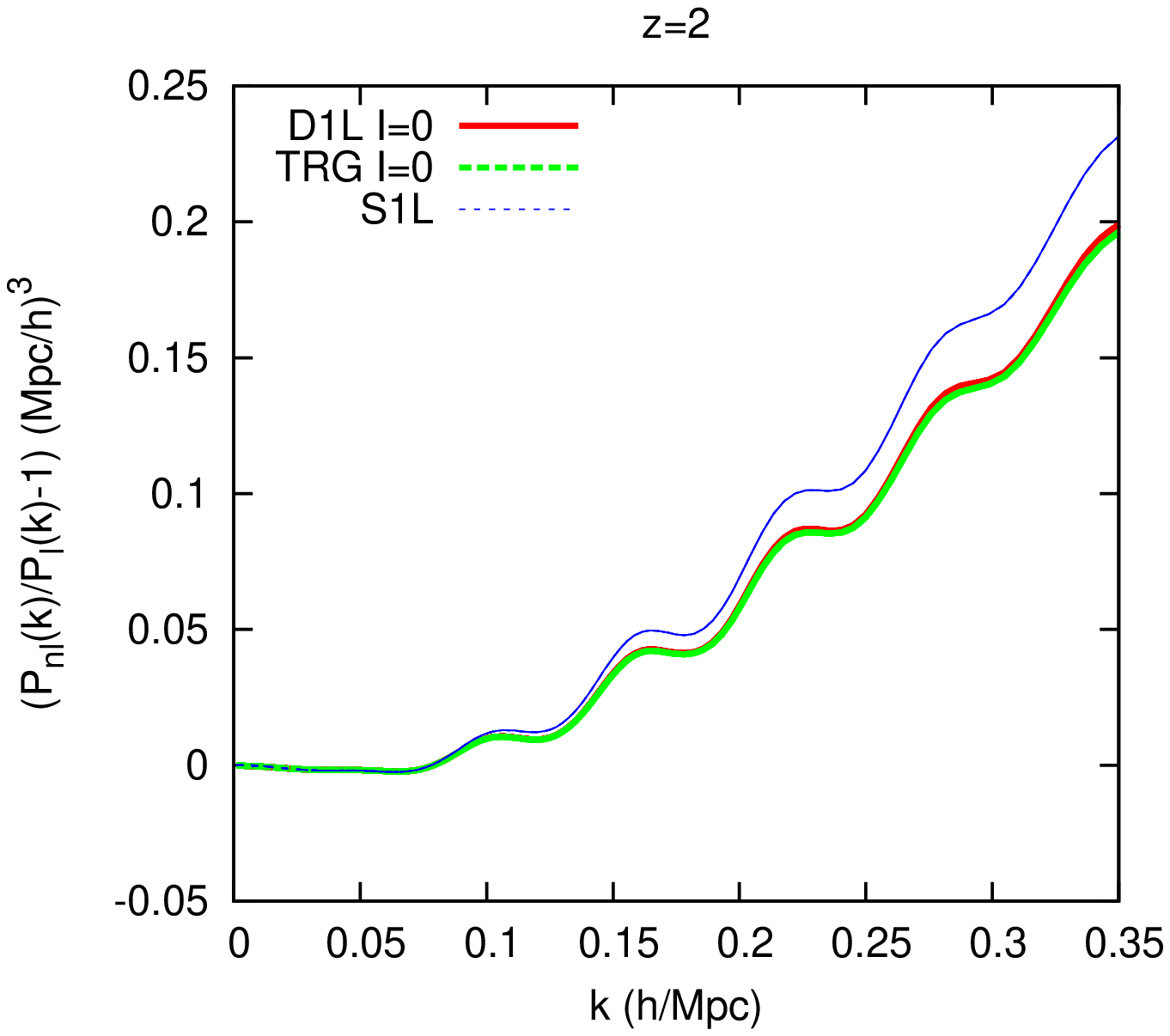}\\
\includegraphics[scale=0.5]{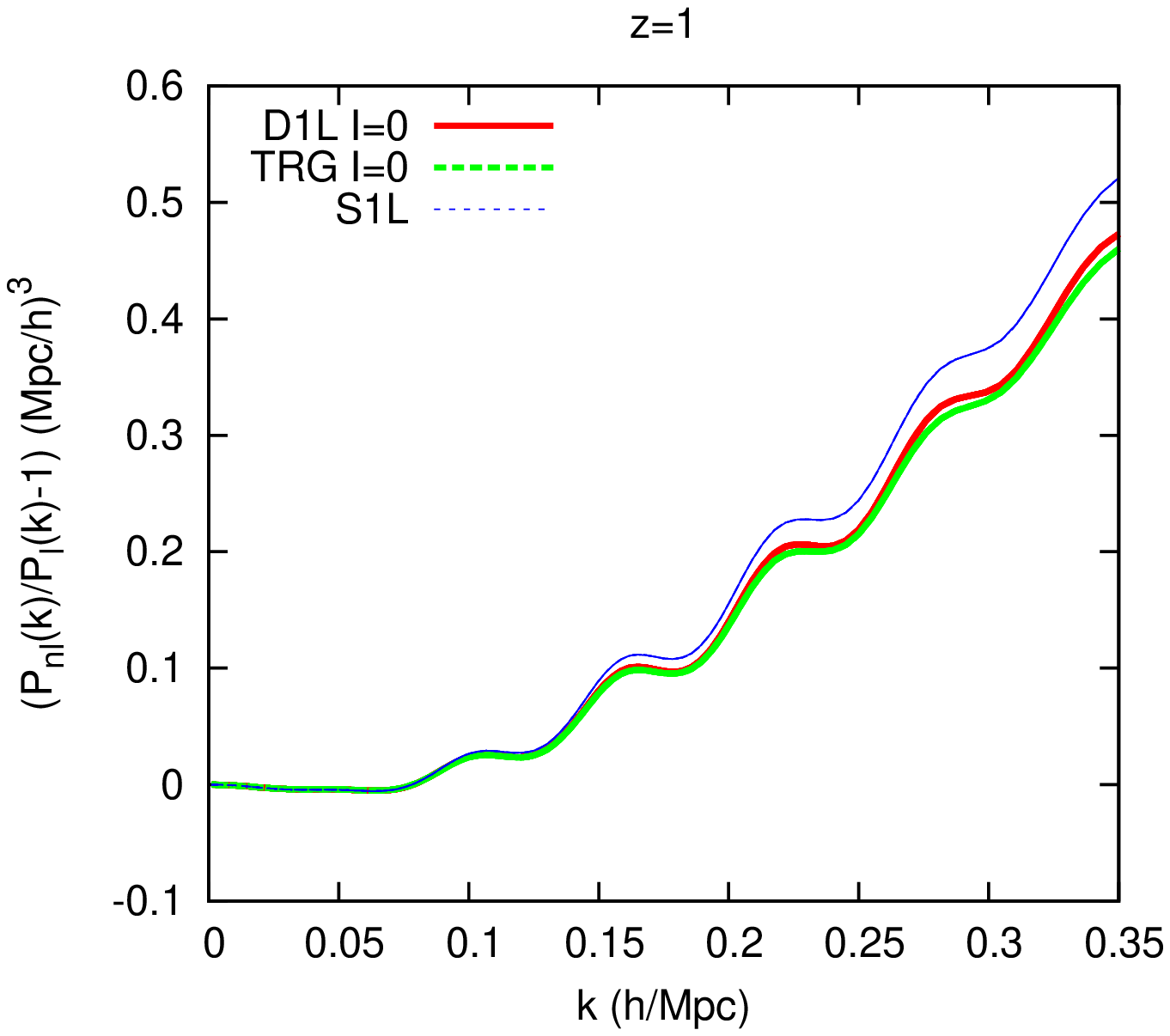}
\includegraphics[scale=0.5]{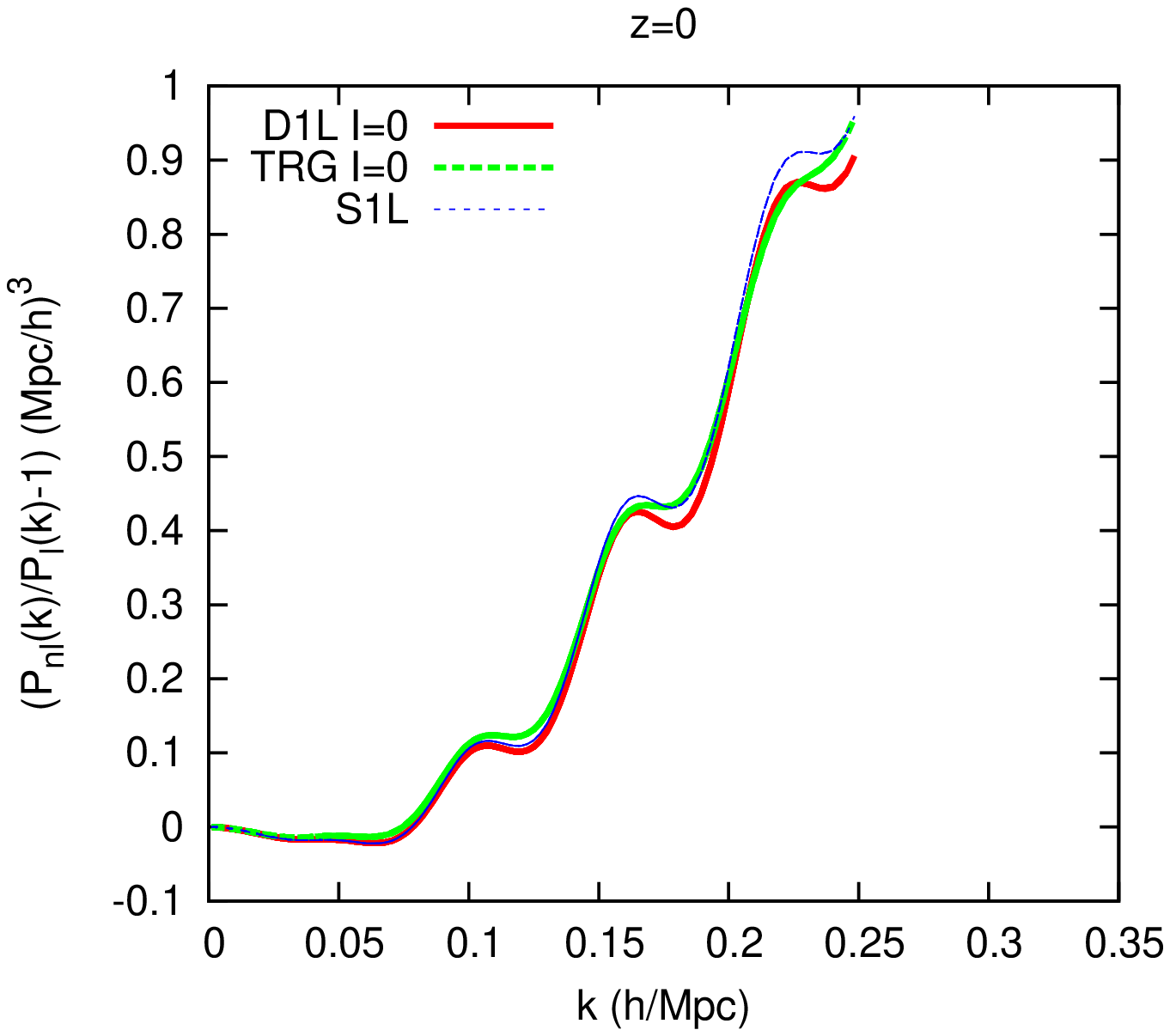}\\
\caption{\label{fig:eds-comp-I0} In an ideal EdS limit and at four
  different redshifts, comparison between the non-linear spectra
  computed with the standard one-loop S1L, dynamical 1-loop D1L and
  TRG methods. For the D1L and TRG case, we start the computation
  from exactly linear initial conditions at $z_{\rm ini}=35$ (with
  vanishing bispectrum), causing a residual difference with respect to
  the S1L method.} }
\FIGURE{
\includegraphics[scale=0.5]{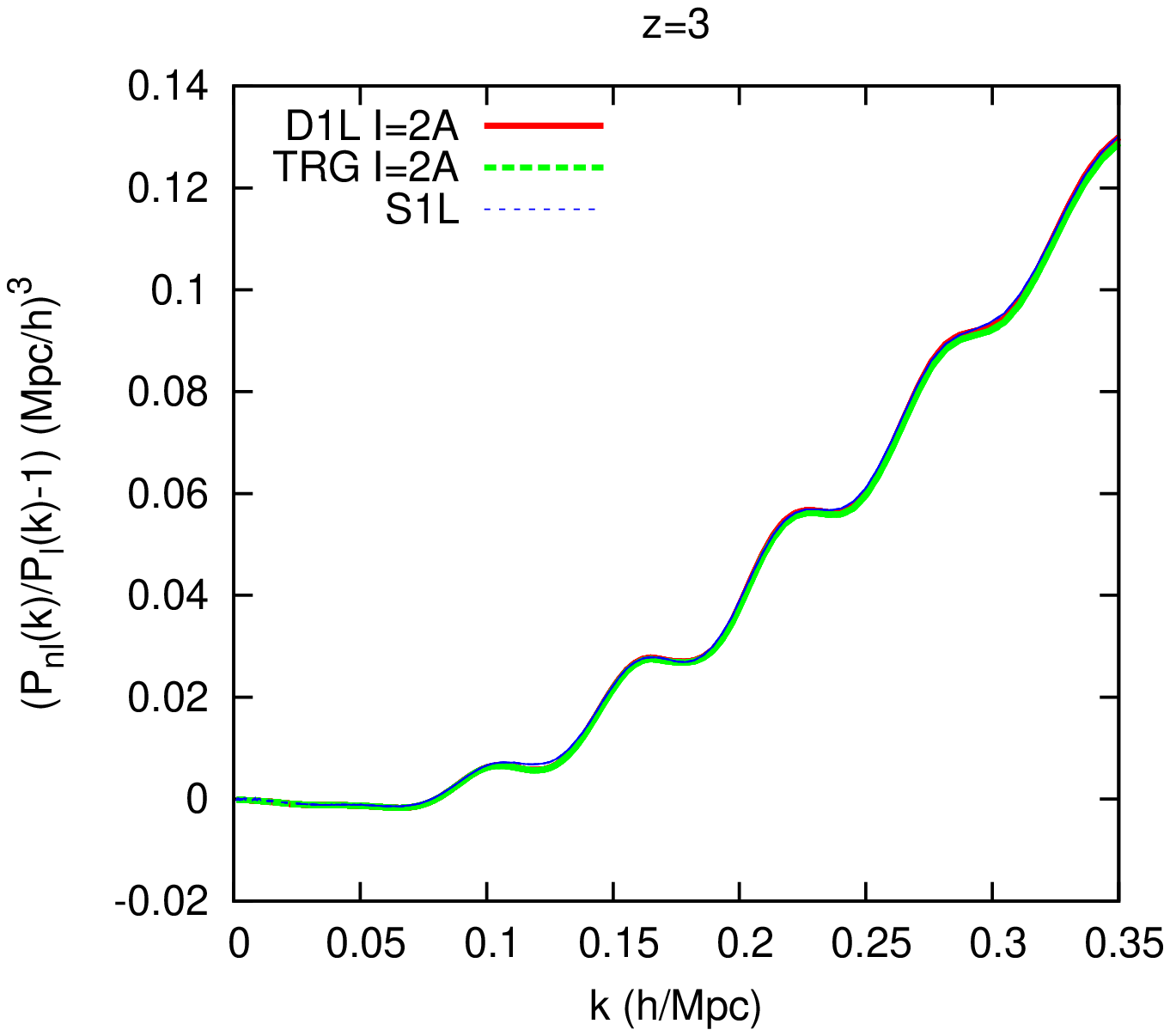}
\includegraphics[scale=0.5]{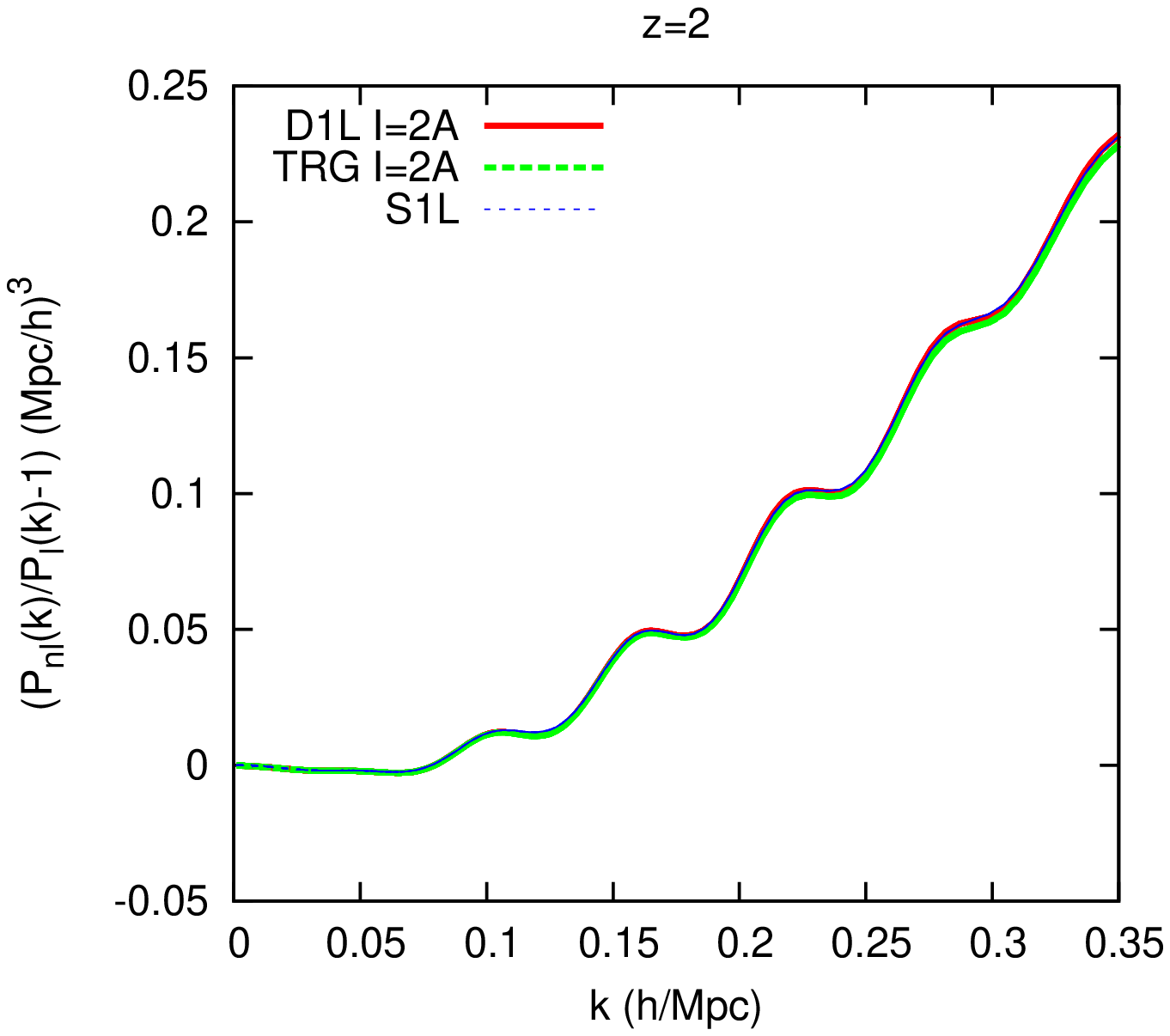}\\
\includegraphics[scale=0.5]{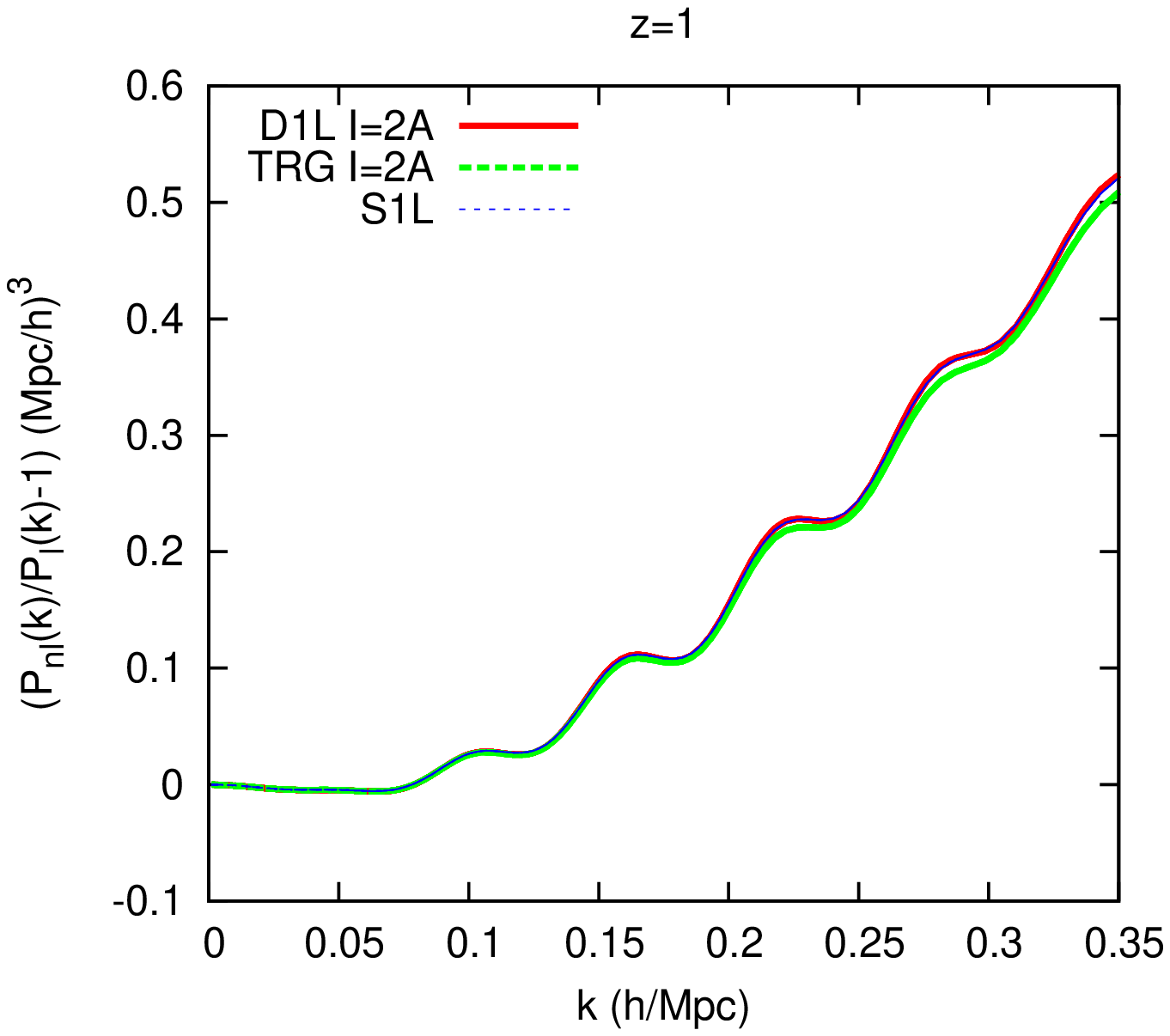}
\includegraphics[scale=0.5]{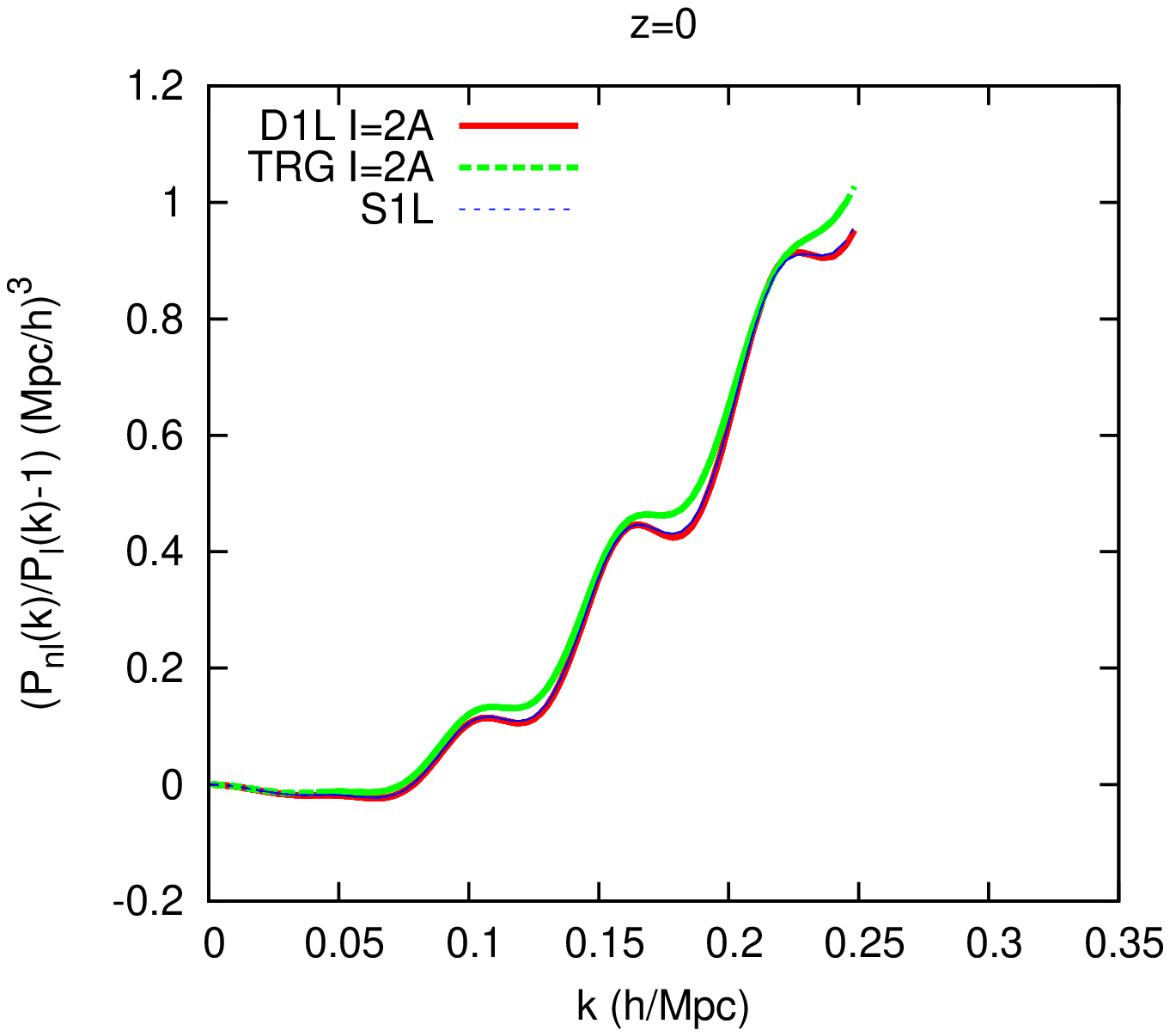}\\
\caption{\label{fig:eds-comp-I2A} Same as previous figure, although
  for the D1L and TRG case we now start from an initial condition
  $I=2A$, which amounts in sharing the same assumptions as the S1L
  calculation.  } 
}

\section{Self-consistency checks in the exact EdS limit\label{sec:eds}}

The main purpose of this section is to test the agreement between the
S1L and D1L methods. This check is more straightforward in a model
without non-trivial growth factors related to $\Omega_m$ begin
smaller than one, i.e. in the Einstein-De Sitter limit.  Hence, for the
results of this section, we modified our TRG/D1L module in order to
enforce a perfect EdS evolution. This amounts in:
\begin{itemize}
\item replacing the actual matrix elements $\Omega_{2b}$ by
  $(3/2,-3/2)$,
\item for D1L, replacing the exact growth functions by $g(\eta)=1$, $f(\eta)=1$.
\end{itemize}
The issue of initial conditions is crucial. The simplest choice
consists in starting from exactly linear perturbations at $z=z_{\rm
  ini}$, i.e. from a $P_{11}(k)$ given by the linear density power
spectrum, from $P_{22}(k)=P_{12}(k)=P_{11}(k)$, and from
$I_{acd,bef}(k)=0$.  Starting from such conditions at $z_{\rm
  ini}=35$, we obtain the 1-loop density power spectra named D1L in
figure~\ref{fig:eds-comp-I0}. They are actually smaller than S1L
results, at all redshift and by several percents (at
$k=0.25\,h$Mpc$^{-1}$ they predict non-linear spectra differing
respectively by $1.2, 1.5, 2, 5\%$ at $z=3,2,1,0$). This discrepancy
actually decreases when we start the integration at a higher
redshift. This suggests that the difference arises from neglecting any
non-linear evolution before $z_{\rm ini}$. Indeed, the S1L method
assumes that from $z\longrightarrow \infty$ till today, the modes have
always evolved according to the newtonian equations in the EdS limit,
with a linear growth factor proportional to $a$. This is of course an approximation, but in order to do a consistent comparison of the methods, we
should also integrate the D1L and TRG equations starting from initial conditions reflecting the
1-loop evolution in the range $z \in [\infty,z_{\rm ini}]$, instead of
linear initial conditions.

This amounts in taking the initial $P_{ab}(k)$ to be the
one-loop spectra at $\eta=0$, and in computing $I_{acd,bef}(k)$ at 
$\eta=0$ at leading order in perturbation theory. However, one-loop
corrections to the linear $P_{ab}(k)$ at $z\sim 35$ appear to be completely
negligible in the range of interest (at least until
$k\sim1\,h$Mpc$^{-1}$), so sticking to the linear spectra makes no
difference in practice (we checked this explicitely). This is not true as far as the initial
bispectra are concerned: instead of $B=0$, we should use the
tree-level bispectrum expression \cite{Bernardeau01}, computed at $z_{\rm ini}$ out of the linear power spectra.
When plugged into the expression of $I_{acd,bef}(k)$, the tree-level bispectrum leads to a series of terms
that we computed explicitely. We checked numerically that all terms are negligible with respect to the leading one, namely $I=2A$: so, for simplicity, we will approximate the tree-level expression of $I$ as $2A$.
A more intuitive derivation of this result follows from observing equation~(\ref{eqn:dIfriendly}). At very early time, the bispectrum is
driven away from zero by the source $A_{acd,bef}$, so that
\begin{equation}
\partial_\eta I_{acd,bef}(k)
\simeq 2 e^{\eta} A_{acd,bef}(k)~.
\end{equation}
Since at leading order in perturbation theory $A_{acd,bef}(k)$ is
constant in time, this approximate evolution equation is solved by
$I_{acd,bef}(k)=2 e^{\eta} A_{acd,bef}(k)$, which corresponds to the
initial condition $I_{acd,bef}(k)=2A_{acd,bef}(k)$ at $\eta=0$.  

When adopting such initial conditions, the D1L results nicely
coincide with S1L predictions, as can be seen in
figure~\ref{fig:eds-comp-I2A}. This was of course expected on a purely
analytical basis, but the agreement proves that the two methods are
correctly implemented numerically.

The most important result of this paper is already visible in Figures~\ref{fig:eds-comp-I0} and~\ref{fig:eds-comp-I2A}: on the scales displayed in
the figures, the TRG curves are always very close to their D1L
counterpart with the same initial conditions, at least for $z \geq
2$. At $z=1$, differences at the level of half-a-percent start to
appear. At $z=0$ one can see clear differences of 1 or 2\% on the
scales corresponding to BAO dips: the TRG smoothes BAOs more than
1-loop approaches. However, this difference between D1L and TRG results
remains small, especially in regard of the large increase in computing
time in the TRG case.  Since in a realistic scenario the growth of
structure is suppressed with respect to the EdS case, this conclusion
is likely to hold even better in $\Lambda$CDM, as we shall see below.

\FIGURE{
\includegraphics[scale=0.5]{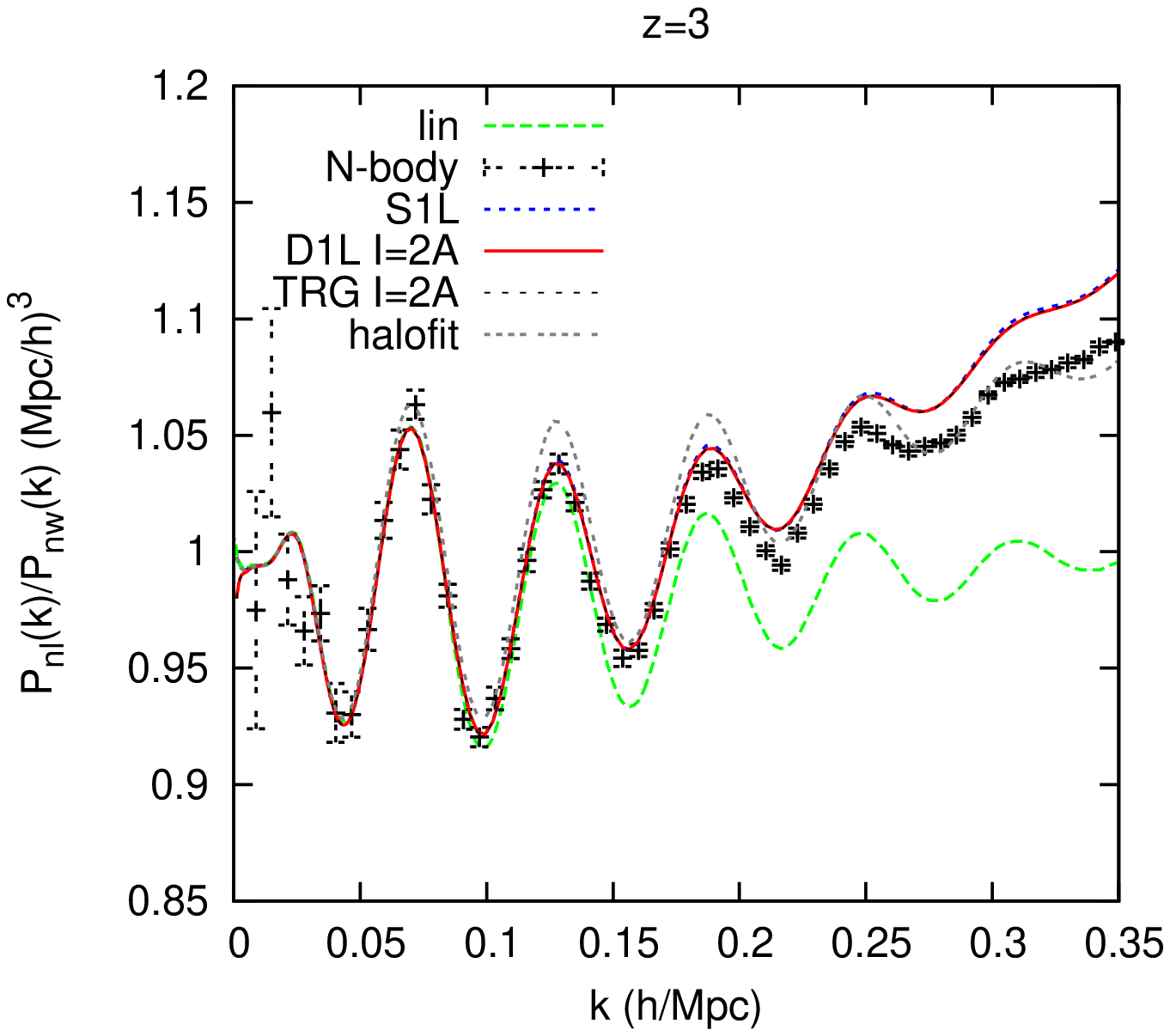}
\includegraphics[scale=0.5]{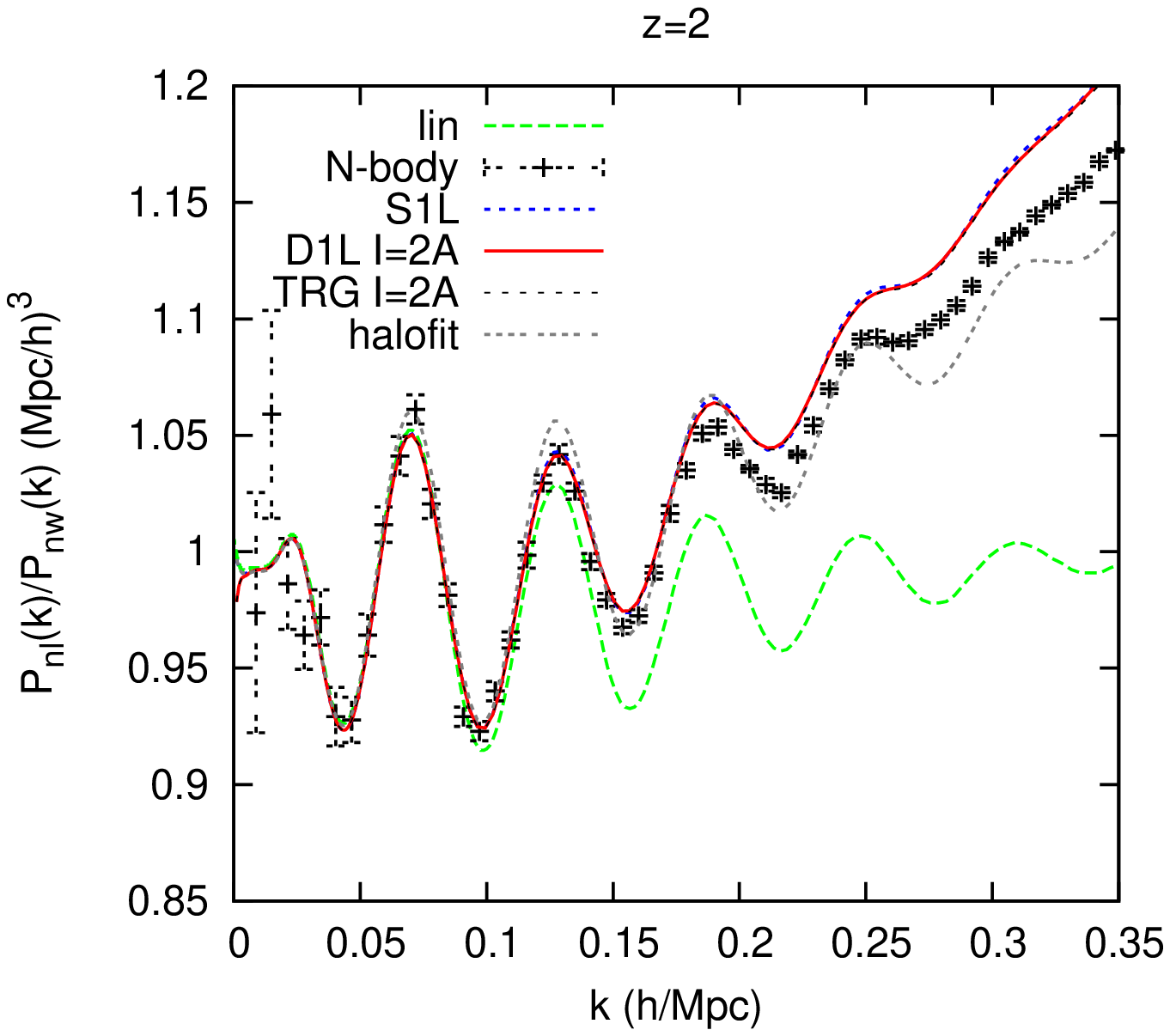}\\
\includegraphics[scale=0.5]{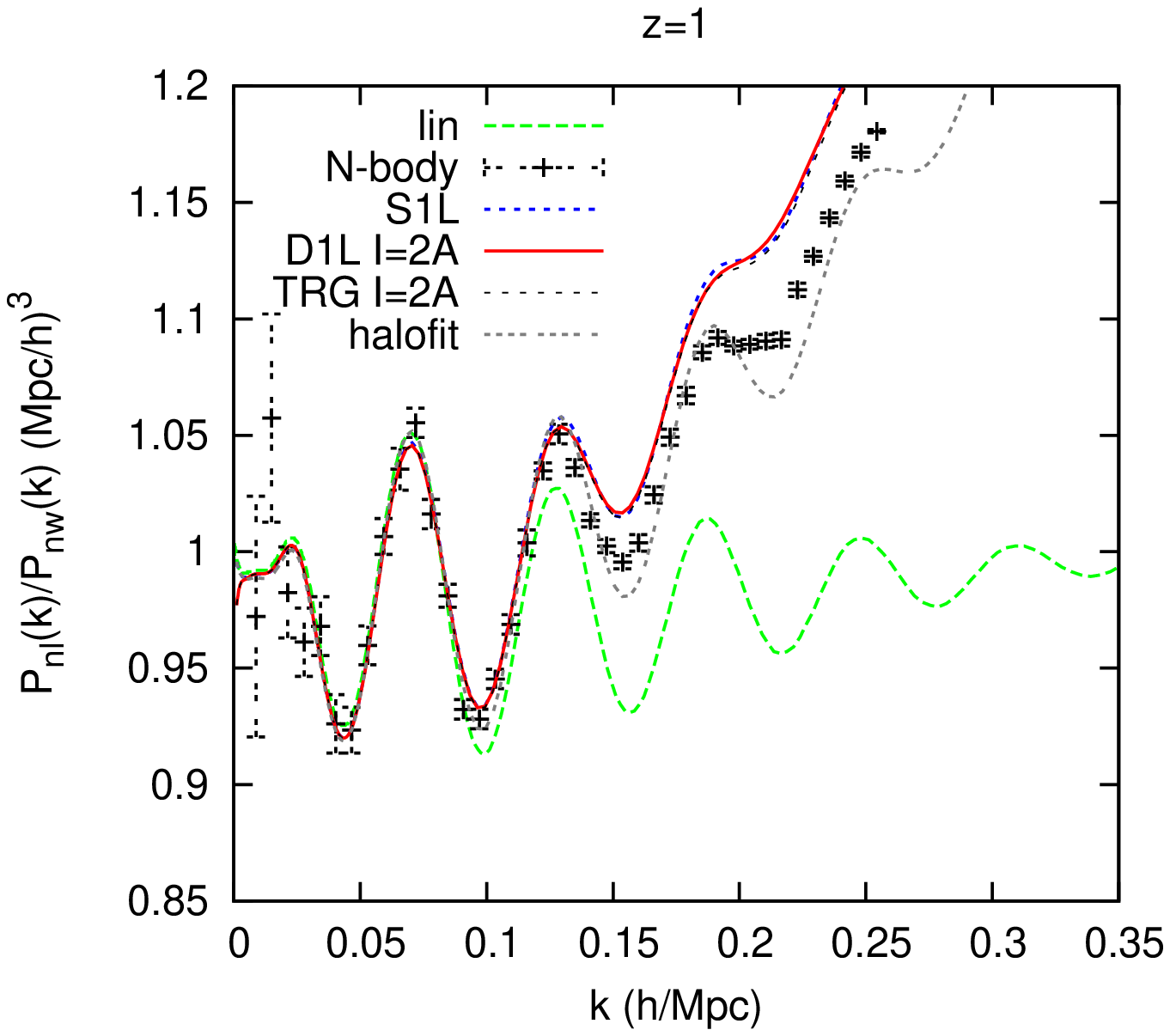}
\includegraphics[scale=0.5]{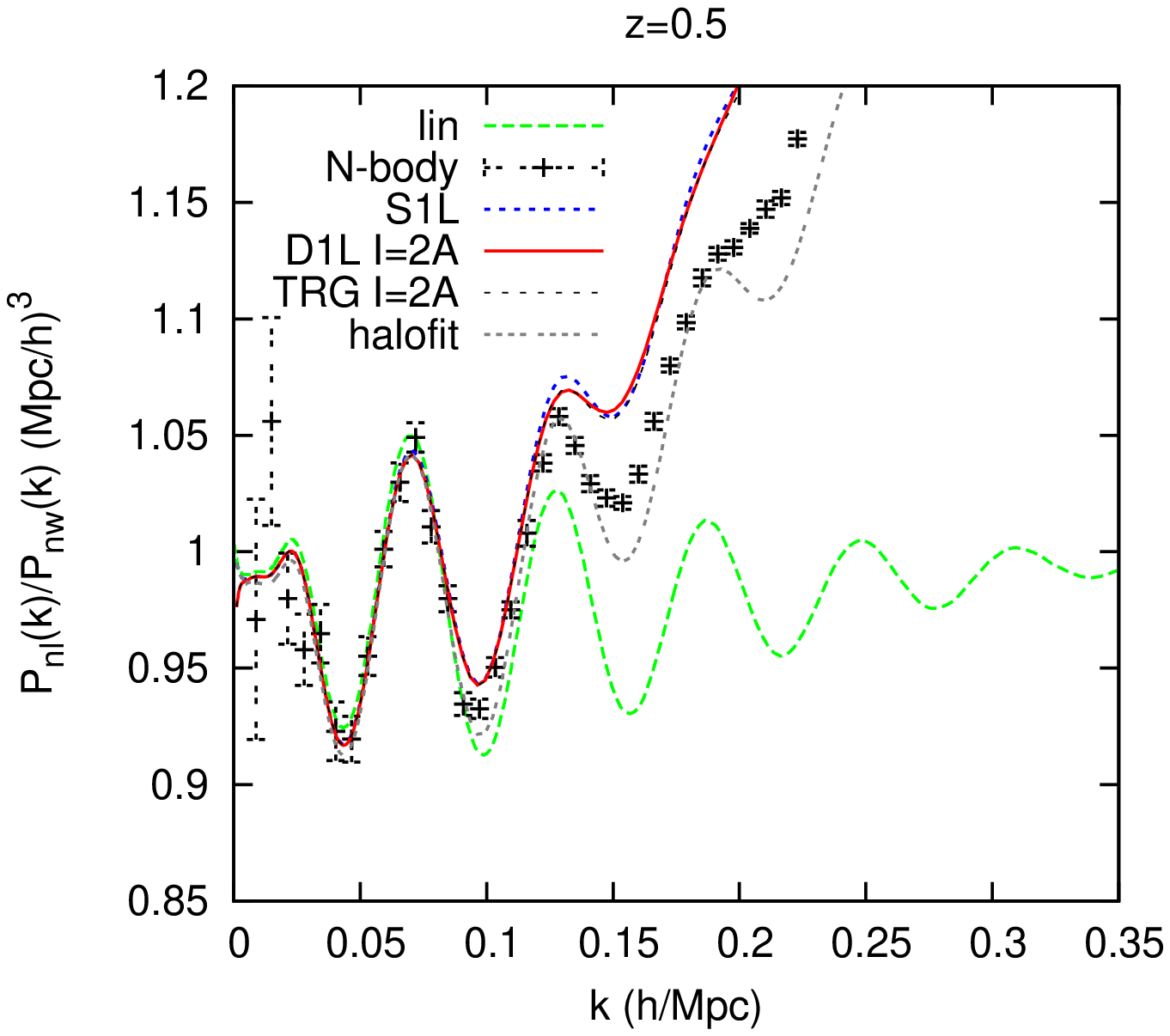}\\
\caption{\label{fig:comp-I2A} Matter power spectrum divided by a smooth
  one~\cite{Eisenstein:1997ik} for $z=3,2,1,0.5$ obtained with five
  different methods: N-body simulations~\cite{Sato:2011qr}, TRG
  method, standard one-loop S1L, dynamical one-loop D1L, and finally
  \HALOFIT{}~\cite{Smith:2002dz}.  Here we employed non-physical
  initial conditions $I=2A$ in order to check the agreement between
  D1L and S1L results, but we know that these results cannot be
  trusted.  }
}

\FIGURE{
\includegraphics[scale=0.5]{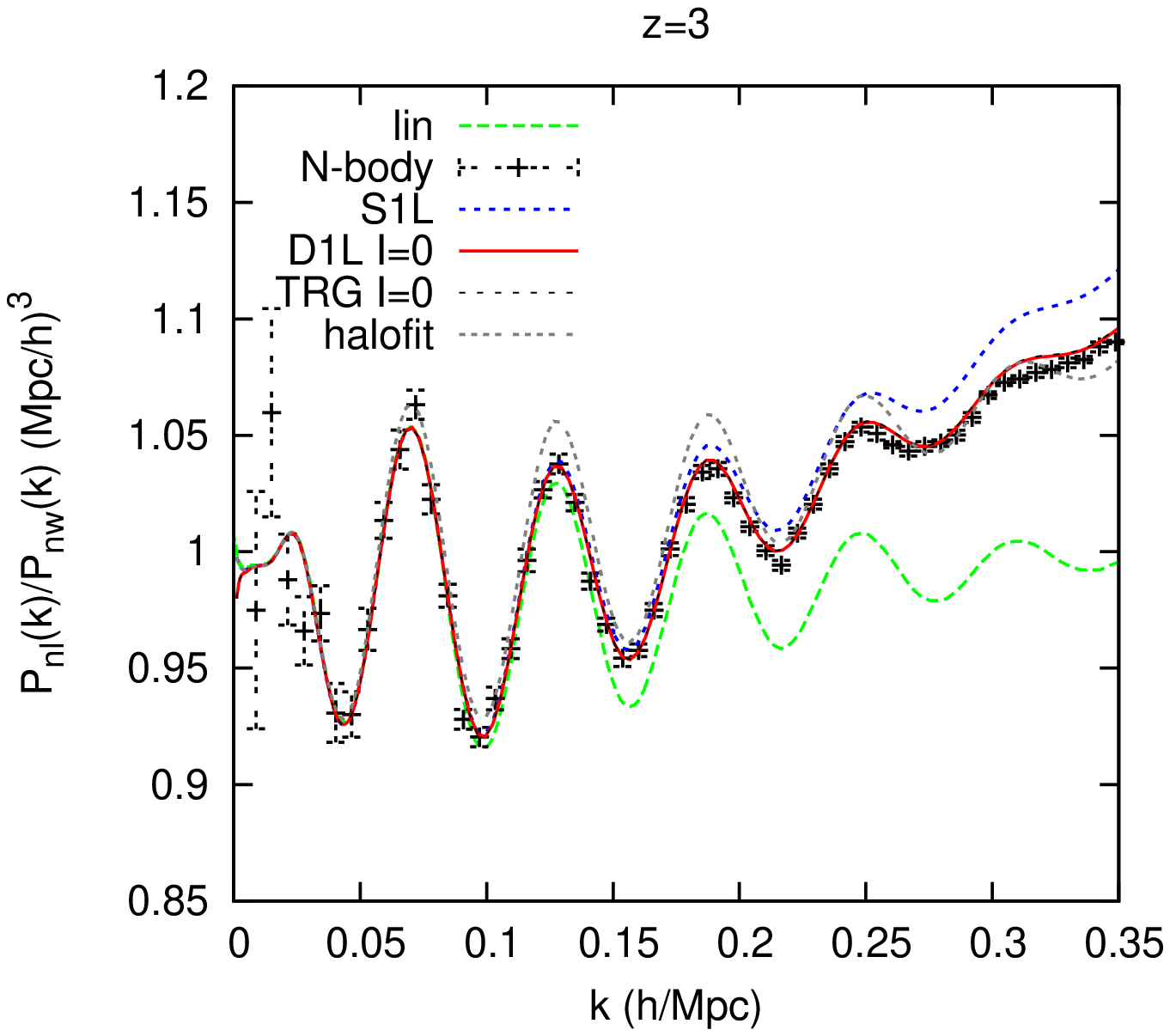}
\includegraphics[scale=0.5]{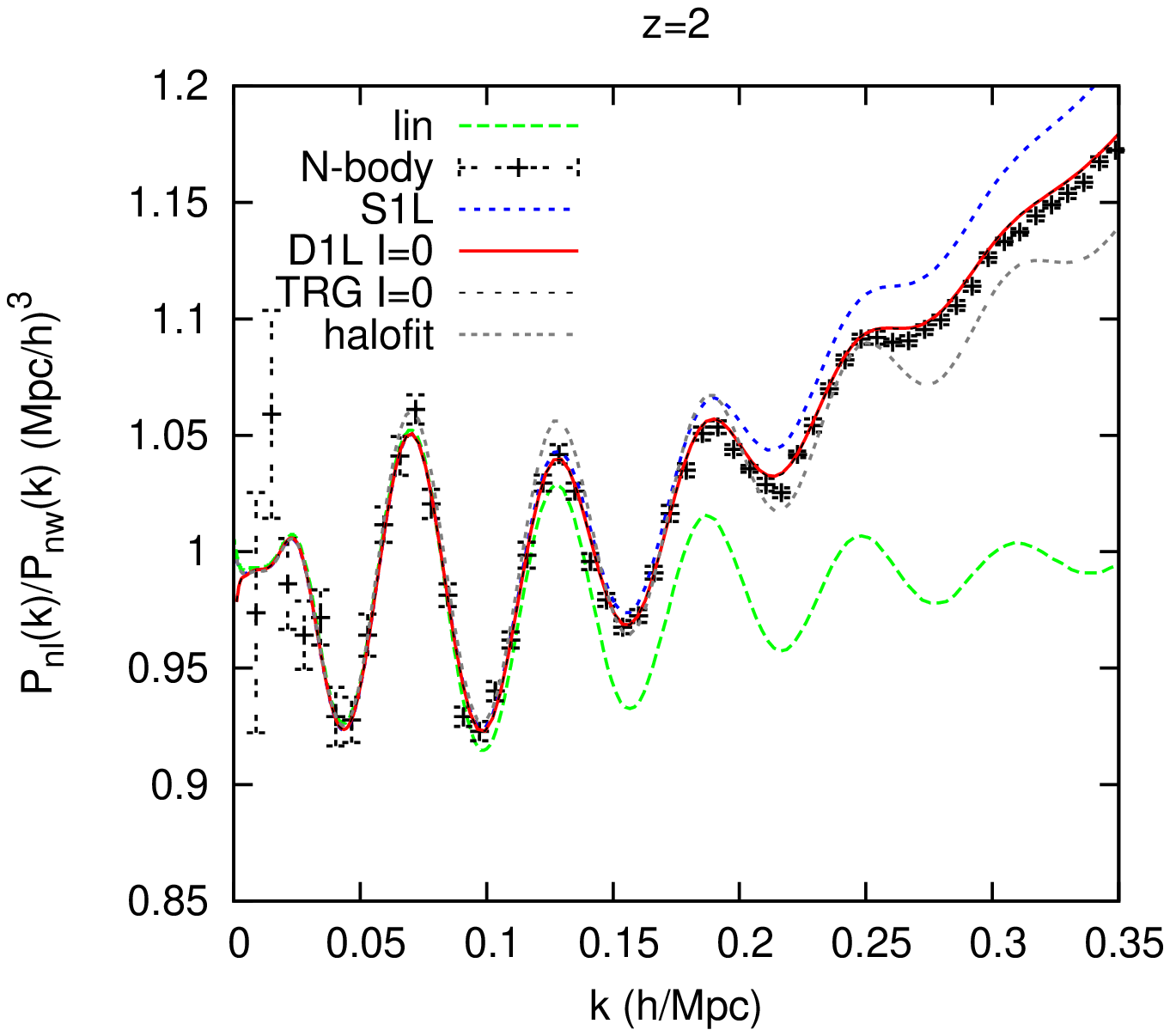}\\
\includegraphics[scale=0.5]{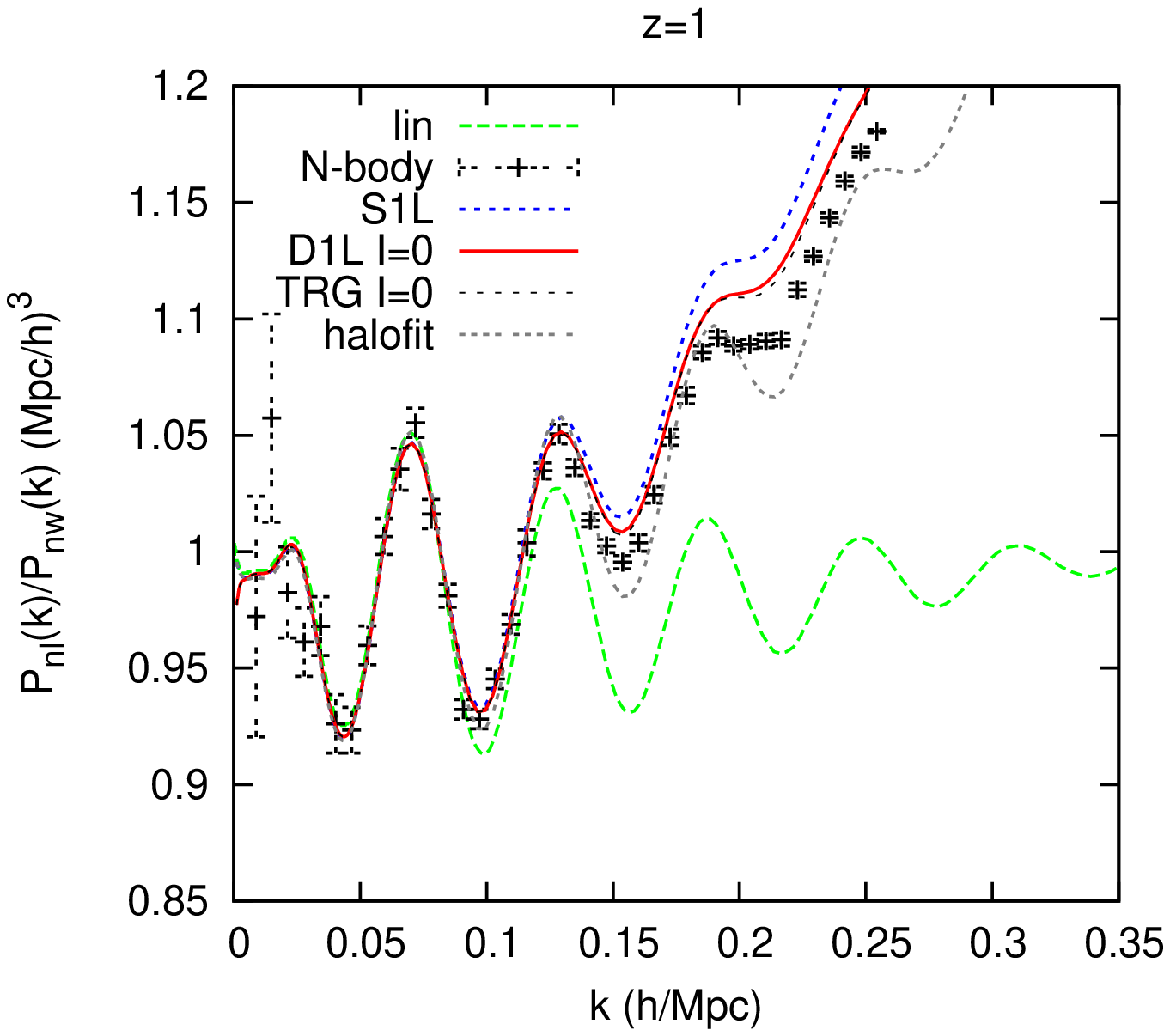}
\includegraphics[scale=0.5]{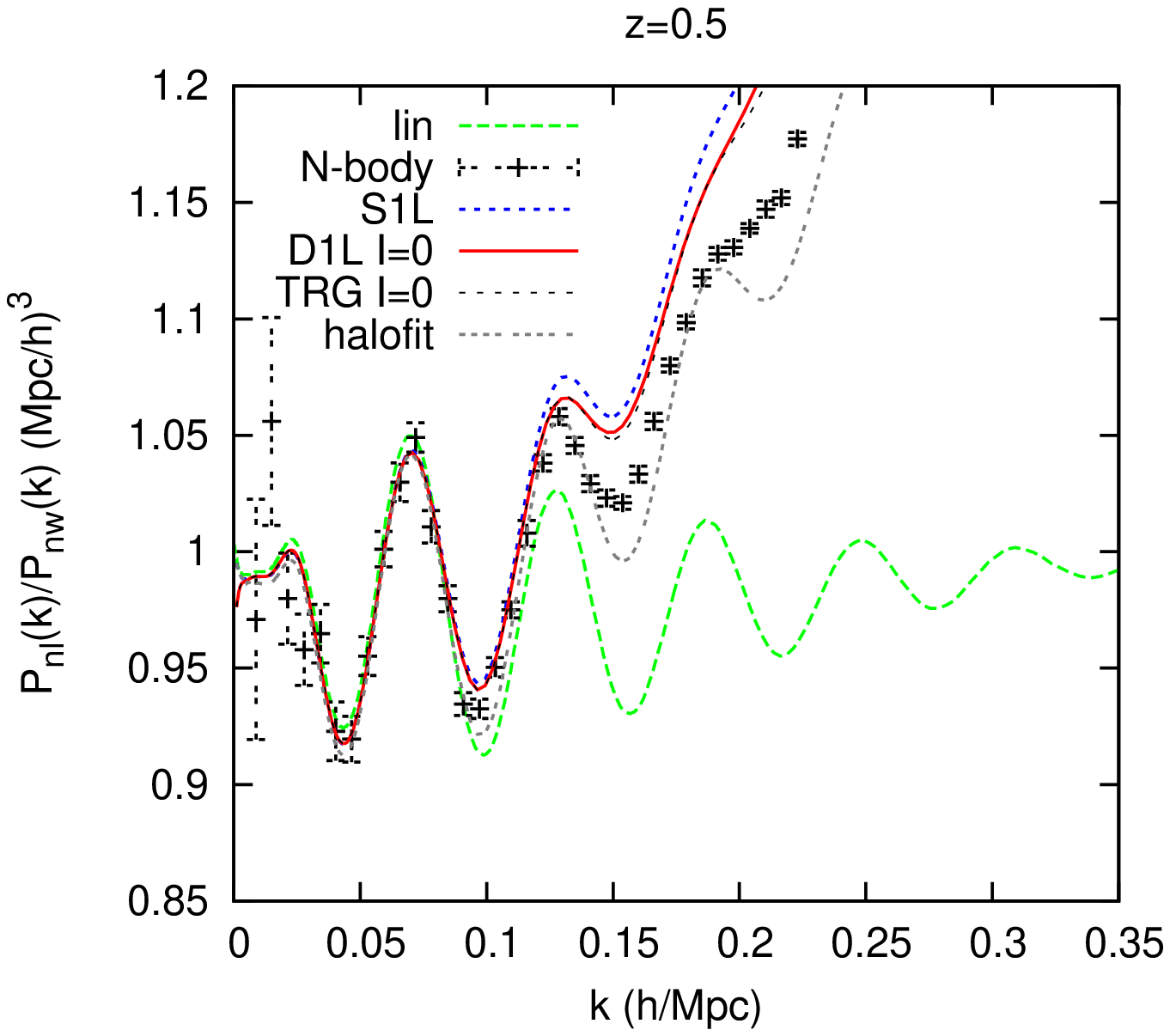}\\
\caption{\label{fig:comp-I0} These figures contain our main results. The matter power spectrum divided by a smooth
  one~\cite{Eisenstein:1997ik} for $z=3,2,1,0.5$ obtained with five
  different methods: N-body simulations~\cite{Sato:2011qr}, TRG
  method, standard one-loop S1L, dynamical one-loop D1L, and finally
  \HALOFIT{}~\cite{Smith:2002dz}. In contrast with the previous figure,
  we now assume linear initial conditions ar $z_{\rm ini}$. This assumption appears to provide excellent results on all displayed scales till $z=2$, and still better predictions than halofit  at $z=1,0$. 
}
}

\section{Results for $\Lambda$CDM and comparison with N-body simulations\label{sec:results}}

We finally ran our code for a  $\Lambda$CDM model, in order to check that the findings of the previous section remain true, namely: {\it (i)} the fact that the D1L method with $I=2A$ matches S1L predictions (despite the fact that D1L treats in a more exact way the impact of $\Omega_m<1$ at early time due to the small radiation density, and the consequence of late $\Lambda$ domination for the time evolution of the density and velocity linear growth factors), and {\it (ii)} the fact that TRG results do not improve significantly over one-loop calculations.

Also, in order to evaluate the precision of each method, we wish to compare
semi-analytic predictions with very accurate N-body simulations, leading
to a density power spectrum with a good resolution in $k$-space, and a
small sampling variance on mildly non-linear scales, hopefully
quantified by an error bar. Such power spectra can only be obtained by
running several simulations in a very large box, and computing the
mean and variance of the resulting collection of spectra.  Simulations
with these characteristics have been presented in several papers
including Refs.~\cite{Crocce:2007dt,Sato:2011qr,Kim:2008kf}. In
Refs.~\cite{Crocce:2007dt,Kim:2008kf}, the initial linear spectrum was
inferred from analytical fitting formulas, which are not accurate at
the per-cent level. We prefer to use the recently published
simulations of Sato et al.~\cite{Sato:2011qr}, based on an initial
spectrum computed with \CAMB{}. These authors performed a large
number of high-quality simulations, leading to sub-percent level
statistical errors, and to a very good matching between linear and
N-body spectra for wavenumbers in the range $[0.05-0.1]h$/Mpc.

We show in figures~\ref{fig:comp-I2A}, \ref{fig:comp-I0} a comparison between
those results and the non-linear spectra obtained with the one-loop
(S1L and D1L), TRG and \HALOFIT{} semi-analytical methods. All spectra
are computed for a model with $h=0.71$, $\Omega_\Lambda=0.735$,
$\Omega_b=0.0448$, $N_{\rm eff} = 3.04$, $T_{\rm cmb}=2.726$K,
$n_s=0.963$ and $\sigma_8=0.80$.

Like in the previous section, we have to face the crucial issue of
initial conditions. We know that starting from the tree-level bispectrum, equivalent to $I=2A$, amounts in doing the same assumption as in the S1L calculation: it
assumes that the bispectrum has grown like in an ever-matter dominated
universe, containing a single presureless species described by Newtonian equations. Actually, most recent N-body simulations (including the one quoted here) also share this assumption. Indeed, their initial conditions
are set by a 2LPT (2nd-order Lagrange Perturbation Theory) algorithm, designed in such way to minimize transient effects. This result is achieved precisely by assuming a tree-level initial bispectrum at initial time. Hence, the choice $I=2A$ is the proper one for comparing D1L or TRG results both with S1L predictions and with N-body results.

We show in figure~\ref{fig:comp-I2A} the D1L and TRG results obtained
from such initial conditions. The S1L and
D1L method are again in very good agreement. We could expect D1L and
S1L to depart from each other at $z\leq1$, when $\Lambda$ domination
leads to non-trivial growth factor which render the expansion of
eq.~(\ref{eq:s1linlambda}) only approximate. This difference can be observed at
$z=0.5$ on the scale $k=0.125\,h$/Mpc corresponding to a maximum in
the BAOs. However, it is only a sub-percent effect, which confirms the
results of Ref.~\cite{Pietroni08}, figure~7, in which a similar test
was presented. As in the previous section, the TRG results are very close to the
S1L ones , even at $z=0.5$. All three methods tend to overestimate non-linear corrections to the density power spectrum: this trend was already well-known for calculations limited to one-loop,
and our result shows that TRG results make roughly the same error. Percent precision is nevertheless achieved until $k\sim 0.2\,h$Mpc$^{-1}$ for $z \geq 2$, or until $k\sim 0.14\,h$Mpc$^{-1}$ at $z=1$.

For comparison, the results obtained when starting from $I=0$, which are presented in
figure~\ref{fig:comp-I0}, seem to be in much better agreement with N-body simulations. 
Note that earlier comparisons shown in
Ref.~\cite{Pietroni08,LesgourguesPietroni09,Carlson:2009it,D'Amico:2011pf,Anselmi:2011ef}
were based on the same assumptions. However, this comparison is not fully self-consistent, since it relies on different choices of initial conditions in the different methods. Hence, the better agreement observed in this plot is likely to be purely coincidental.

\section{Discussion\label{sec:discussion}}

In this work, we compared a few semi-analytical methods to the recent
accurate N-body simulations of Sato et al.~\cite{Sato:2011qr} and to
standard 1-loop perturbation theory. We released these methods in the
form of a C module, named {\tt trg.c} and integrated in the Boltzmann
code \CLASS{}, version 1.2. The first method is the Time
Renormalization Group (TRG) proposed by M.~Pietroni. The second
method, named here D1L (Dynamical 1-Loop) and based on the same
approach, leads to nearly identical results on scales of interest,
with a much smaller computing time.  We have explained in this paper
our approach for dealing with numerical instabilities, and for
optimizing the algorithm. We have also presented some convergence
tests proving its reliability. This release allows any cosmologist to
compute easily an approximate non-linear spectrum (from first
principles, rather than with fitting formulas), while until now this
task was only accessible to a few specialists.

Differences between TRG and D1L are visible at low redshift,
but they remain very small on scales of interest. This shows that
the partial resumation of diagrams beyond one loop in the TRG method 
improves one-loop results by a negligible amount, for a much larger computing time. Hence, the one-loop limit of the TRG equations (namely, the D1L scheme) is preferable in practice.

The D1L algorithm is a fast and practical tool for obtaining one-loop results for the density/velocity power spectra (and tree-level results for the bispectra), even for non-trivial cosmological models or in presence of an arbitrary initial bispectrum. There are several ways to incorporate such assumptions in one-loop calculations, but in the D1L  equations, this flexibility is present from the beginning.
In this paper, we used this opportunity for showing the importance of assumptions concerning the initial bispectrum. In our universe, the non-linearity of the gravitational equations is sufficient to induce a tiny bispectrum even at very high redshift on cosmological scales of interest in this paper.
At any given initial redshift, this bispectrum is small enough to be well approximated by a tree-level calculation, but nevertheless large enough to impact results at small redshift by several percents. 
This observation is consistent with
previous studies of initial conditions for N-body simulations.  In
particular, the 2LPT method has been developed in order to deal with
transient effects in simulations \cite{Crocce:2006ve}, i.e. in order to remove
spurious decaying modes by implementing initial conditions infered from second-order Lagrangian
perturbation theory, including an initial tree-level bispectrum.
However, this fact had been overlooked in the context of TRG calculations.
We have shown that previous claims that TRG results improve over one-loop predictions were the consequence of neglecting this initial bispectrum. When using the same initial bispectrum in all methods, the difference between S1L,  D1L and TRG results almost disappears.

It is beyond the scope of this paper to discuss to which extent the initial conditions assumed in the S1L method and in N-body simulations offer a sufficiently realistic description of the actual universe. The fact that at high redshift baryons, cold dark matter and radiation perturbations do not share the same transfer functions and do not obey exactly to the growing mode solution means that in principle, 
one should implement some amount of ``physical transient effects'' in the initial conditions of both N-body simulations and renormalisation algorithms. The authors of \cite{Somogyi:2009mh} have already shown the importance of treating baryons separately from cdm when performing non-linear calculations. To our knowledge, the impact of early baryon and radiation perturbations on the calculation of the initial bispectrum (to be passed to N-body 
initial condition generators) has not been quantified accurately. We speculate that the D1L algorithm described in this paper could be a convenient tool for computing realistic high-redshift bispectra, and we leave this for future studies.

\section*{Acknowledgments}

We would like to thank F. Bernardeau, M.~Crocce, M.~Pietroni and R.~Scoccimarro for invaluable comments and
suggestions concerning this manuscript, as well as S.~Anselmi,
G.~Ballesteros, M.~Shaposhnikov and O.~Ruchaysky for very useful
exchanges. We are also very grateful to M.~Sato and T.~Matsubara for
letting us use results from their simulations.  This project is
supported by a research grant from the Swiss National Science
Foundation.

\appendix

\section{From Boltzmann to continuity and Euler equation \label{sec:boltz}}

Let's start by considering a large set of particles only interacting through gravitation. In order to lighten slightly this presentation, let us omit the $m$ subscript for the moment. For each particle (CDM or baryon) the equation of motion is given in terms of its proper time $t$, velocity $\mathbf{v}$ and position $\mathbf{r}$:
\begin{align}
  \frac{\text{d}\mathbf{v}}{\text{d}t}= - \frac{\partial \phi }{\partial \mathbf{r}},
  \label{eqn:start}
\end{align}
where $\phi$ is the Newtonian potential induced by the local mass density $\rho(\mathbf{r})$,
\begin{align}
  \phi(\mathbf{r}) = G \int \text{d}^3\mathbf{r}'\frac{\rho(\mathbf{r}')}{|\mathbf{r}'-\mathbf{r}|}.
  \label{eqn:newtonianpot}
\end{align}
This set of particles is however sitting in an expanding universe. The
gravitational collapse consists in studying departures from the
homogeneous Hubble expansion. It is then useful to redefine the
variables of position and momentum with respect to the comoving
ones. The physical space coordinates $\mathbf{r}$ relate to the
comoving space coordinates $\mathbf{x}$ through $\mathbf{r}\equiv
a(\tau)\mathbf{x}$. We can define the density perturbation
$\delta(\xtau)$, the peculiar velocity $\mathbf{u}(\xtau)$ and the
cosmological gravitational potential $\Phi(\xtau)$ by
%
%
\begin{align}
  \rho(\xtau)&\equiv\bar{\rho}(\tau)[1+\delta(\xtau)]\label{eqn:defdelta},\\
  \mathbf{v}(\xtau)&\equiv\Hub\mathbf{x}+\mathbf{u}(\xtau)\label{eqn:defu},\\
  \mathbf{\phi}(\xtau)&\equiv-\frac12\frac{\partial \Hub}{\partial \tau}\mathbf{x}^2+\Phi(\xtau)\label{eqn:defphi}.
\end{align}
The Einstein equations lead in the sub-Hubble, weak gravitational
field limit to the Poisson equation:
\begin{align}
  \nabla^2\Phi(\xtau)&=-\frac32\Hub^2\sum_i\Omega_i(\tau)\delta_i(\xtau).
  \label{eqn:poisson}
\end{align}
Defining the momentum $\mathbf{p}=am\mathbf{u}$, eq.~(\ref{eqn:start}) now reads:
\begin{align}
  \frac{\text{d}\mathbf{p}}{\text{d}\tau}=-am\nabla\Phi(\mathbf{x}).
  \label{eqn:motion}
\end{align}
One finally write the collisionless Boltzmann equation for the
phase-space density $f(\xptau)$ in its full form:
\begin{align}
  \frac{\text{d} f}{\text{d} \tau}&=\frac{\partial f}{\partial \tau}+ \frac{\mathbf{p}}{ma}\cdot \nabla f - am\nabla \Phi\cdot\frac{\partial f}{\partial \mathbf{p}}.
  \label{eqn:boltzfull}
\end{align}
The non-linearity of this equation is induced by gravitational
back-reaction, described by the Poisson equation. This equation is
obviously hard to solve due to its high dimensionality, which can
however be reduced by taking the momenta of the distribution
$f(\xptau)$, and by truncating such an expansion at some order, if
this can be physically justified. Let us define the first three
momenta as:
\begin{alignat}{2}
  &\int \text{d}^3\mathbf{p}~ f(\xptau)~&\equiv&~ \rho(\mathbf{x},\tau),\\
  &\int \text{d}^3\mathbf{p}~ \frac{\mathbf{p}}{am} f(\xptau)~&\equiv&~ \rho(\xtau)\mathbf{u}(\xtau),\\
  &\int \text{d}^3\mathbf{p}~ \frac{p_ip_j}{a^2m^2} f(\xptau)~ &\equiv&~ \rho(\xtau)u_iu_j(\xtau) + \sigma_{ij}(\xtau).
\end{alignat}
Taking the zeroth moment of eq.~(\ref{eqn:boltzfull}) leads to the continuity equation:
\begin{align}
  \frac{\partial \delta(\xtau)}{\partial \tau}+\nabla\cdot{[1+\delta(\xtau)]\mathbf{u}(\xtau)}=0,\label{continuity}
\end{align}
while the first moment gives (after subtracting $\mathbf{u}(\xtau)$
times (\ref{continuity})) the Euler equation:
\begin{align}
  \frac{\partial u_i(\xtau)}{\partial \tau} + \hub u_i(\xtau) + \left(\mathbf{u}(\xtau)\cdot\nabla\mathbf{u}(\xtau)\right)_i = - \nabla_i \Phi(\xtau)-\frac{1}{\rho}\nabla_j\sigma_{ij}.\label{eqn:euler}
\end{align}
One usually takes the anisotropic pressure $\sigma_{ij}$ to be equal
to zero. This approximation is excellent not just for baryons (they
interact, so they are considered as a perfect fluid with no shear),
but also for cold dark matter. Indeed, CDM particles have no
interactions but a tiny velocity dispersion, leading to negligible
anisotropic pressure on scales where CDM can be represented by a
single-flow distribution. In the range of linear and quasi-linear
scales considered here, this approximation is valid.

With such considerations, we can safely use from now on the $\delta_m$
notation for both baryons and CDM. The Poisson equation
(\ref{eqn:poisson}) can be cast into a more convenient form:
\begin{align}
  \nabla^2\Phi(\xtau)&=-\frac32\Hub^2\Omega_m(\tau)\delta_m(\xtau)\left(1+\sum_{i\neq b,c}\frac{\Omega_i\delta_i}{\Omega_m\delta_m}\right),
  \label{eqn:poissonbetter}
\end{align}
where in $i$ stands for photons and neutrinos (relativistic or not).
It can be seen from eq.~(\ref{eqn:euler}) that the curl part of the
velocity field decouples and is suppressed by the universe expansion.
The equations can then be reformulated in terms of the density
perturbation and the velocity divergence
$\theta=\nabla\cdot\mathbf{u}$. Furthermore, to have a better
understanding of the mode coupling induced by the non-linearity,
we transform the equation into Fourier space, with the following
convention:
\begin{align}
  A(\ktau)&=\frac{1}{(2\pi)^3}\int \text{d}^3\mathbf{x}e^{-i\mathbf{k}\cdot\mathbf{x}}A(\xtau).
  \label{eqn:fourierdef}
\end{align}
In Fourier space, the equations governing the matter field reduce to 
\begin{align}
  \frac{\partial \delta_m(\ktau)}{\partial \tau} + \theta_m(\ktau) &=-\int\text{d}^3\mathbf{q}\text{d}^3\mathbf{p}\delta_D(\mathbf{k}-\mathbf{p}-\mathbf{q})\alpha(\pq)\theta_m(\ptau)\delta_m(\qtau),\label{eqn:contF}\\
  \frac{\partial \theta_m(\ktau)}{\partial \tau}+\hub\theta_m(\ktau)&\nonumber\\
+\frac32\hub^2\sum_i\Omega_i(\tau)\delta_i(\ktau) &= - \int\text{d}^3\mathbf{q}\text{d}^3\mathbf{p}\delta_D(\mathbf{k}-\mathbf{p}-\mathbf{q})\beta(\pq)\theta_m(\ptau)\theta_m(\qtau),\label{eqn:eulerF}
\end{align}
where the kernels $\alpha$ and $\beta$ encoding the mode coupling are defined as:
\begin{align}
  \alpha(\pq)&\equiv \frac{(\mathbf{p}+\mathbf{q})\cdot\mathbf{p}}{p^2},\hspace{2cm}  \beta(\pq) \equiv \frac{(\mathbf{p}+\mathbf{q})^2(\mathbf{p}\cdot\mathbf{q})}{2p^2q^2}.\label{eqn:alphabeta}
\end{align}
\section{Structure of the {\tt trg.c} module in \CLASS\label{sec:class}}

As explained in \cite{Lesgourgues:2011re}, \CLASS{} is organized in
eleven modules. The role of the ninth module {\tt nonlinear.c} is to
evaluate the non-linear power spectra according to the method chosen
by the user. In the current version \CLASS{} v1.1, the {\tt nonlinear.c}
module can optionally call the {\tt trg.c} module described in this
paper, in one out of three modes ({\tt non linear} = {\tt test-linear},
{\tt one-loop} or {\tt trg}) already described in subsection
\ref{sec:3modes}), and with a set of precision parameters.  After the
execution of the {\tt trg.c} module, the code is ready to write in
output files the non-linear density, velocity and cross spectra at
different redshifts chosen by the user.

The {\tt trg.c} module consists of several functions.  The most
important one, \verb+trg_init()+, is called from the {\tt nonlinear.c}
module.  Its goal is to compute the non-linear power spectrum from
a given starting redshift and linear power spectrum previously computed
by the {\tt spectra.c} module. 

The \verb+trg_init()+ routine first defines relevant step sizes in
$\eta$ space and in $k$ space. For the latter, it calls the two functions
\verb+trg_logstep1_k()+ and \verb+trg_logstep2_k()+. The $k$ steps are
defined in such way to keep a high resolution in the
baryon-oscillation zone, and a reasonably small number of steps on
other scales. Since these steps are relevant for the double escape
procedure defined previously, one might want to be cautious while
modifying them.

The initial density spectrum $P_{11}$ is taken directly from the {\tt
  spectra.c} module.  The initial velocity and cross spectra $P_{22}$
and $P_{12}$ are computed by evaluating a finite difference between
$P_{11}$ at $\eta_{ini} \pm \epsilon$.  At the starting redshift,
three-points correlating functions (bispectra) are assumed to be zero,
so all $I$'s defined in eq.~(\ref{eqn:defI}) are initially zero.

The integration of the TRG equations (full equations in {\tt trg}
mode, simplified equations in the other two modes) is then performed.
At each step in the {\tt trg} mode, or only once in the {\tt one-loop}
mode, \verb+trg_init()+ calls \verb+trg_integrate_xy_at_eta+ to
perform explicitly the integrals in $(x,y)$ space and find each of the
14 $A$ factors.  This last function evaluates each of the 14
integrands at each $(x,y)$ point by calling the function
\verb+trg_A_arg_trg()+ (or \verb+trg_A_arg_one_loop()+).

After the execution of the {\tt trg.c} module, the non-linear spectra
are stored in a structure {\tt nonlinear} associated with the {\tt
  nonlinear.c} module. They can be evaluated at any value of $k$ and
$z$ by calling the function {\tt nonlinear\_pk\_at\_k\_and\_z()}. Instead,
all values at a given $z$ are returned by the function {\tt
  nonlinear\_pk\_at\_z()}. The output module {\tt output.c} actually
calls this last function in order to write the final results in
separate files for the density, velocity and cross spectra.

\bibliographystyle{utcaps}
\bibliography{trg}

\providecommand{\href}[2]{#2}\begingroup\raggedright\begin{thebibliography}{10}

\bibitem{Hannestad:2002cn}
S.~Hannestad, ``{Can cosmology detect hierarchical neutrino masses?},'' {\em
  Phys.Rev.} {\bf D67} (2003) 085017,
  \href{http://arXiv.org/abs/astro-ph/0211106}{{\tt astro-ph/0211106}}.

\bibitem{Smith:2002dz}
{\bf The Virgo Consortium} Collaboration, R.~Smith {\em et al.}, ``{Stable
  clustering, the halo model and nonlinear cosmological power spectra},'' {\em
  Mon.Not.Roy.Astron.Soc.} {\bf 341} (2003) 1311,
  \href{http://arXiv.org/abs/astro-ph/0207664}{{\tt astro-ph/0207664}}.

\bibitem{Bernardeau01}
F.~Bernardeau, S.~Colombi, E.~Gaztanaga, and R.~Scoccimarro, ``{Large-scale
  structure of the universe and cosmological perturbation theory},'' {\em Phys.
  Rept.} {\bf 367} (2002) 1--248,
\href{http://arXiv.org/abs/astro-ph/0112551}{{\tt astro-ph/0112551}}.

\bibitem{Crocce:2005xy}
M.~Crocce and R.~Scoccimarro, ``{Renormalized cosmological perturbation
  theory},'' {\em Phys.Rev.} {\bf D73} (2006) 063519,
  \href{http://arXiv.org/abs/astro-ph/0509418}{{\tt astro-ph/0509418}}.

\bibitem{Crocce:2007dt}
M.~Crocce and R.~Scoccimarro, ``{Nonlinear Evolution of Baryon Acoustic
  Oscillations},'' {\em Phys.Rev.} {\bf D77} (2008) 023533,
  \href{http://arXiv.org/abs/0704.2783}{{\tt 0704.2783}}.

\bibitem{Pietroni08}
M.~Pietroni, ``{Flowing with Time: a New Approach to Nonlinear Cosmological
  Perturbations},'' {\em JCAP} {\bf 0810} (2008) 036,
\href{http://arXiv.org/abs/0806.0971}{{\tt 0806.0971}}.

\bibitem{Carlson:2009it}
J.~Carlson, M.~White, and N.~Padmanabhan, ``{A critical look at cosmological
  perturbation theory techniques},'' {\em Phys.Rev.} {\bf D80} (2009) 043531,
  \href{http://arXiv.org/abs/0905.0479}{{\tt 0905.0479}}.

\bibitem{Sato:2011qr}
M.~Sato and T.~Matsubara, ``{Nonlinear Biasing and Redshift-Space Distortions
  in Lagrangian Resummation Theory and N-body Simulations},''
\href{http://arXiv.org/abs/1105.5007}{{\tt 1105.5007}}.

\bibitem{Lesgourgues:2011re}
J.~Lesgourgues, ``{The Cosmic Linear Anisotropy Solving System (CLASS) I:
  Overview},'' \href{http://arXiv.org/abs/1104.2932}{{\tt 1104.2932}}.

\bibitem{Blas:2011rf}
D.~Blas, J.~Lesgourgues, and T.~Tram, ``{The Cosmic Linear Anisotropy Solving
  System (CLASS) II: Approximation schemes},''
  \href{http://arXiv.org/abs/1104.2933}{{\tt 1104.2933}}.

\bibitem{Springel:2005mi}
V.~Springel, ``{The Cosmological simulation code GADGET-2},'' {\em
  Mon.Not.Roy.Astron.Soc.} {\bf 364} (2005) 1105--1134,
  \href{http://arXiv.org/abs/astro-ph/0505010}{{\tt astro-ph/0505010}}.

\bibitem{LesgourguesPietroni09}
J.~Lesgourgues, S.~Matarrese, M.~Pietroni, and A.~Riotto, ``{Non-linear Power
  Spectrum including Massive Neutrinos: the Time-RG Flow Approach},'' {\em
  JCAP} {\bf 0906} (2009) 017,
\href{http://arXiv.org/abs/0901.4550}{{\tt 0901.4550}}.

\bibitem{D'Amico:2011pf}
G.~D'Amico and E.~Sefusatti, ``{The nonlinear power spectrum in clustering
  quintessence cosmologies},'' \href{http://arXiv.org/abs/1106.0314}{{\tt
  1106.0314}}.

\bibitem{Anselmi:2011ef}
S.~Anselmi, G.~Ballesteros, and M.~Pietroni, ``{Non-linear dark energy
  clustering},'' \href{http://arXiv.org/abs/1106.0834}{{\tt 1106.0834}}.

\bibitem{Eisenstein:1997ik}
D.~J. Eisenstein and W.~Hu, ``{Baryonic features in the matter transfer
  function},'' {\em Astrophys.J.} {\bf 496} (1998) 605,
  \href{http://arXiv.org/abs/astro-ph/9709112}{{\tt astro-ph/9709112}}.

\bibitem{Kim:2008kf}
J.~Kim, C.~Park, I.~Gott, Richard, and J.~Dubinski, ``{The Horizon Run N-body
  Simulation: Baryon Acoustic Oscillations and Topology of Large Scale
  Structure of the Universe},'' {\em Astrophys.J.} {\bf 701} (2009) 1547--1559,
  \href{http://arXiv.org/abs/0812.1392}{{\tt 0812.1392}}.

\bibitem{Crocce:2006ve}
M.~Crocce, S.~Pueblas, and R.~Scoccimarro, ``{Transients from Initial
  Conditions in Cosmological Simulations},'' {\em Mon.Not.Roy.Astron.Soc.} {\bf
  373} (2006) 369--381, \href{http://arXiv.org/abs/astro-ph/0606505}{{\tt
  astro-ph/0606505}}.

\bibitem{Somogyi:2009mh}
G.~Somogyi and R.~E. Smith, ``{Cosmological perturbation theory for baryons and
  dark matter I: one-loop corrections in the RPT framework},'' {\em Phys.Rev.}
  {\bf D81} (2010) 023524, \href{http://arXiv.org/abs/0910.5220}{{\tt
  0910.5220}}.

\end{thebibliography}\endgroup

\end{document}